\documentclass[aps,pra,reprint,superscriptaddress,showpacs]{revtex4-1}

\usepackage{graphicx}
\usepackage{color}
\usepackage{braket}
\usepackage{amsmath}
\usepackage[version=3]{mhchem}

\usepackage{epstopdf}

\usepackage{lipsum}
\usepackage[T2A]{fontenc}
\usepackage[utf8]{inputenc}
\usepackage[russian,english]{babel}

\usepackage{comment}

\begin{document}
\title{Number squeezed and fragmented states of strongly interacting bosons in a double well}
\author{Joel C.~Corbo}
\email[Email: ]{joel.corbo@colorado.edu}
\altaffiliation[Current affiliation: ]{Center for STEM Learning, University of Colorado Boulder, Boulder, CO 80309}
\affiliation{Berkeley Center for Quantum Information and Computation, University of California, Berkeley, CA 94720, USA}
\affiliation{Department of Physics, University of California, Berkeley, CA 94720, USA}
\author{Jonathan L.~DuBois}
\affiliation{Lawrence Livermore National Lab, 7000 East Ave, L-415, Livermore, CA 94550, USA}
\author{K.~Birgitta Whaley}
\affiliation{Berkeley Center for Quantum Information and Computation, University of California, Berkeley, CA 94720, USA}
\affiliation{Department of Chemistry, University of California, Berkeley, CA 94720, USA}


\begin{abstract}
We present a systematic study of the phenomena of number squeezing and fragmentation for a
repulsive Bose-Einstein condensate (BEC) in a three dimensional double well potential over a range
of interaction strengths and barrier heights, including geometries that exhibit appreciable
overlap in the one-body wavefunctions localized in the left and right wells.
We compute the properties of the condensate with numerically exact, full dimensional
path integral ground state (PIGS) Quantum Monte Carlo simulations and compare with results
obtained from using two- and eight-mode truncated basis models. The truncated basis models
are found to agree with the numerically exact PIGS simulations for weak interactions,
but fail to correctly predict the amount of number squeezing and fragmentation exhibited
by the PIGS simulations for strong interactions. We find that both number squeezing and fragmentation
of the BEC show non-monotonic behavior at large values of interaction strength $a$.
The number squeezing shows a universal scaling with the product of number of particles and interaction
strength ($Na$) but no such universal behavior is found for fragmentation. Detailed analysis
shows that the introduction of repulsive interactions not only suppresses number fluctuations
to enhance number squeezing, but can also enhance delocalization across wells and tunneling
between wells, each of which may suppress number squeezing. This results in a dynamical competition
whose resolution shows a complex dependence on all three physical parameters defining the system:
interaction strength, number of particles, and barrier height.
\end{abstract}

\date{\today}
\maketitle

\section{Introduction}
\label{sec:intro}

Since the achievement of Bose-Einstein condensation in the laboratory~\cite{Davis1995,Anderson1995},
there has been significant experimental and theoretical interest in the study of a
Bose-Einstein Condensate (BEC) in
a double well trap. This system has been modeled extensively using several different
theoretical approaches.  One common approach is to use a variant of the Bose-Hubbard model
within an $n$-mode approximation, in which the many-body state of the system is computed in terms of
a basis constructed from the system's $n$ lowest energy one-body states.  The bulk of this work
has been done in the context of a two-mode
model~\cite{Mahmud2002,Mahmud2005,Milburn1997,Javanainen1999,Bodet2010,Spekkens1999,Mueller2006,Sakmann2011,Ananikian2006,Jia2008,Luhmann2012,Zapata1998,Meier2001},
although some authors have gone beyond two-mode models by including four~\cite{Gillet2014}
or eight modes~\cite{Garcia-March2010}.  Other authors have studied multi-mode effects
with multiconfigurational Hartree-Fock methods, in which a basis is constructed from
a generalization of the two-mode basis using bosonic Hartree-Fock
theory~\cite{Masiello2005,Streltsov2006,Zollner2006,Sakmann2008}, Gross-Pitaevski mean field
methods~\cite{Ostrovskaya2000,Sinatra2000,Li2009,Kurkjian2017}, or a semiclassical truncated Wigner
approximation~\cite{Isella2006,Tuchman2009,Gross2011}.

As discussed in~\cite{Javanainen1999}, the two-mode model is generally restricted to the limit of
weak interactions. Most studies focus on models which include only one-body tunneling ($J$)
and on-site interaction ($U$) terms, although some include higher-order effects either through
two-body tunneling terms~\cite{Ananikian2006,Mahmud2005,Jia2008} or renormalized $J$ and $U$
coefficients~\cite{Luhmann2012}.  Recent theoretical work on understanding
the double well in the context of an eight-mode approximation~\cite{Garcia-March2010}
has uncovered some deficiencies of the two-mode model.  In particular, it was shown that even
in regimes where one might naively expect only the lowest two one-body modes
to contribute based on energy arguments, the ground state can nevertheless contain components
of higher modes. Such admixtures have the potential to dramatically influence the collective
properties of the system in ways that are not captured by including two-body effects
within a two-mode model.

One such property is the degree of \emph{number squeezing} exhibited by the system's many-body
wavefunction (from now on, when we refer to \emph{squeezing} in this paper, we mean number squeezing). Number squeezing is related to the probability of finding particular values for
the difference in the number of particles on the two sides of the double well. In an unsqueezed system,
the probabilities are distributed
as a classical Gaussian centered on a difference of zero; thus, while the most probable configuration
is the one with an equal number of particles on each side of the barrier, there is non-negligible
probability of finding other configurations.  A system is said to be squeezed when this distribution
narrows, resulting in non-classical distributions for which the probability of observing
a difference of zero dominates over all other possibilities. The extreme limit of such symmetric
squeezing is the specific ``number correlated''~\cite{Dunningham2002} state wiith equal numbers
of particles in each well, sometimes referred to as a dual or ``twin'' Fock
state~\cite{Holland1993,Xiang2010,Pezze2013}. Number squeezing can be introduced by increasing
the strength of the repulsive inter-particle interactions (e.g., via a Feshbach
resonance~\cite{Feshbach1958,Pollack2009}) or by decreasing the tunneling strength between the wells
(e.g., by increasing the barrier height~\cite{Gross2011}).

In recent years, a number of experimental studies have realized number squeezing of BECs in harmonic
traps~\cite{Chuu2005,Appel2009}, double well potentials~\cite{Sebby2007,Jo2007,Gross2010}
and optical lattice systems~\cite{Orzel2001,Greiner2002a,Gerbier2006,Li2007,Esteve2008,Itah2010}.
Number squeezed states are important for atom-optics applications, in particular for interferometry
and the use of interferometry for precision measurements, sensing, and metrology~\cite{Cronin2009}.
The microscopic coherence of BECs makes them attractive candidates for atom interferometry based
on either interference between internal hyperfine states~\cite{Riedel2010} or interference between
spatially separated condensates~\cite{Shin2004,Schumm2005,Berrada2013}.  In the latter situation,
a single trapped BEC is first split into two clouds by introducing a double well potential,
and then the phase between the spatially separated components is subsequently measured
from the interference fringes that result from the overlap of the components after ballistic
expansion~\cite{Andrews1997,Shin2004}.

BECs with non-classical number correlations show further advantages for interferometry over
classical condensates, possessing extended phase diffusion
times~\cite{Gati2006a,Li2007,Sebby2007,Jo2007,Folling2007,Ferrini2008}, greater robustness
to atom loss~\cite{Dunningham2002,Dunningham2004}, and greater sensitivity.   In particular,
the extreme ``twin Fock'' number squeezed states have been shown to be capable of phase sensing
below the standard quantum limit of $N^{-1/2}$, where $N$ is the total number of particles
involved in the measurement; in principle these states
can even achieve the Heisenberg limit of $N^{-1}$~\cite{Pezze2013}. Phase sensitivity can also
be significantly enhanced by squeezing the collective pseudo-spin of the two-mode
system~\cite{Wineland1992,Kitagawa1993}, given by the ratio of the number squeezing
and its conjugate variable, the relative phase, rather than just the relative number
fluctuations~\cite{Bollinger1996,Hald1999,Kuzmich2000,Gross2010,Gross2011,Leroux2010,Leroux2010a,Riedel2010,Chen2011,Louchet2010,Gross2014}.
These properties make both number and spin squeezed condensates desirable starting points
for designing BEC interferometers to realize interferometric measurements with precision
scaling below the standard quantum limit. The role of non-classical number squeezed states
in enhancing the coherence time was recently demonstrated in a full Mach-Zehnder interferometer
for BECs in tunable double well potentials integrated on an atom chip~\cite{Berrada2013},
while a number of experiments have demonstrated significant spin squeezing in BECs trapped
in double well or optical lattice
potentials\cite{Gross2010,Gross2011,Leroux2010,Riedel2010,Chen2011,Gross2014}.

The symmetry of the double well potential also allows for
\emph{fragmentation} of the condensate~\cite{Nozieres1982}. In a canonical BEC, we can express
the system's many-body wavefunction in such a way that all particles are in the same one-body state.
However, under certain conditions the BEC can exhibit fragmentation, in which multiple
one-body states are macroscopically occupied by the particles in the condensate.
When population is further distributed over one-body states with non-macroscopic occupation,
the BEC is said to be \emph{depleted}.  Experimental studies have confirmed
the presence of depletion in an atomic BEC in an optical lattice~\cite{Xu2006} as well as
fragmentation in a quasi-1D atomic BEC in a magnetic waveguide~\cite{Leanhardt2002}.

Theoretical analysis of number squeezing has largely been made with two-mode models for double well
systems~\cite{Imamoglu1997,Villain1997,Javanainen1999,Spekkens1999,Ferrini2008,Averin2008,Bodet2010,Sakmann2011,Julia2012}
and with Bose-Hubbard analogs of the two-mode model for optical lattice
systems~\cite{Burnett2002,Capogrosso2007}. No analysis of number squeezing has yet been made
within the eight-mode model or with full three-dimensional computational simulations,
although a one-dimensional truncated Wigner approximation has been used to estimate
on-site~\cite{Isella2006} and inter-site number fluctuations~\cite{Gross2011}. Regarding
fragmentation, much theoretical effort has also gone into understanding this phenomenon using
two-mode models~\cite{Spekkens1999,Streltsov2004,Mueller2006,Jackson2008,Bader2009,Sakmann2011}.
As is the case with squeezing, these models are inadequate to fully capture the behavior
of the system with respect to fragmentation at large interaction strengths. Indeed, studies that
employed multiconfigurational Hartree-Fock methods have already demonstrated evidence
of  fragmentation behavior beyond that which is seen in traditional two-mode
models~\cite{Streltsov2006,Masiello2005}. Like squeezing, neither fragmentation nor depletion
has yet been studied within the eight-mode model.

In this paper, we develop a more complete understanding of the phenomena of number squeezing
and fragmentation for a repulsive Bose-Einstein condensate in a three dimensional double well
potential, by making a systematic study over a wide range of interaction strengths and
barrier heights and comparing results from finite basis models with results from numerically exact
Quantum Monte Carlo (QMC) calculations.  We first conduct the analysis of these properties
in the context of the two- and eight-mode models, extending the scope of these models
as required. For this analysis, we use a basis of states constructed from one-body,
non-interacting states of the double well potential, which is the basis typically employed
in previous work on double well systems.  Our analysis provides new information within
the restricted domain of these models.  In particular, we explicitly examine well geometries
that allow overlap between one-body wave functions localized in left and right wells,
and show that this overlap gives rise to a non-monotonic dependence of both squeezing
and fragmentation on interaction strength.

We then employ the full many-body formalism of QMC to evaluate the ground state properties
of the BEC using the path integral ground state (PIGS)
method~\cite{Sarsa2000,Moroni2004,Cuervo2005,Krauth1996}. The numerically
exact PIGS method allows us to move beyond the range of validity of the $n$-mode models
and into the regime of strongly interacting systems in three-dimensions within a continuum
representation. We determine the amount of squeezing and fragmentation present in the BEC
as a function of interaction strength, finding a marked non-monotonic dependence of these
on the interaction strength for higher barriers. We also find that number squeezing shows
a universal scaling on the product $Na$, where $N$ is the total number of particles
and $a$ is the $s$-wave scattering length that parameterizes the inter-particle interaction;
this scaling makes our results applicable to systems with a wide range of particle number $N$.
We then compare the results of these numerically exact calculations with the predictions
of the two- and eight-mode models; in the two-mode case, we compare to both calculations made
with the non-interacting basis and calculations made with a basis constructed from solutions
to the Gross-Pitaevskii equation.  We find that none of these models correctly predict
the squeezing and fragmentation behavior of the system for sufficiently large $Na$,
but their comparison leads nevertheless to an understanding of when, how, and why
these approximate models break down.  In particular, the non-monotonic behavior of number squeezing
at large $Na$ values is found to be a result of the competing effects of interactions
and delocalization afforded by higher energy states, showing that a multi-mode picture
is essential for understanding squeezing as either the interaction strength or the number
of particles (or both) increases.

\section{The System}

\subsection{The Many-Body Double Well Hamiltonian}
\label{sec:hamiltonian}

\begin{figure}
\includegraphics{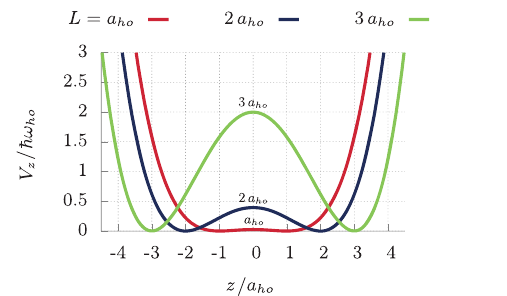}
\caption{(Color online) The $z$ component of the external potential for
$\alpha = 4/81\, a_{ho}^{-2}$. The height of the barrier,
$V_{ext}(0) = m\omega_{ho}^{2}\alpha L^{4}/2$, is $2/81\,\hbar\omega_{ho}$,
$32/81\,\hbar\omega_{ho}$, and $2\,\hbar\omega_{ho}$
for $L=a_{ho}$, $2\, a_{ho}$, and $3\, a_{ho}$, respectively.}
\label{fig:dwell}
\end{figure}

The many-body Hamiltonian for $N$ bosons of mass $m$ interacting pairwise in an external potential
has the form
\begin{equation}
\hat{H} = \sum_{j=1}^{N} \left( -\frac{\hbar^2}{2m} \nabla_{j}^{2} + V_{ext}(\mathbf{r}_{j}) \right) + \sum_{j<k}^{N} V_{int}(\mathbf{r}_{j},\mathbf{r}_{k}).
\end{equation}
For the external potential, we use a three-dimensional double well potential of the form
\begin{equation}
V_{ext}(\mathbf{r}) = \frac{1}{2}m\omega_{ho}^{2} \left(x^{2} + y^{2} + \alpha\left(z^{2}-L^{2}\right)^{2} \right),
\label{eq:double_well_pot}
\end{equation}
where $\omega_{ho}$ is the characteristic harmonic trap frequency in the $xy$ plane,
$\alpha$ characterizes the height of the barrier between wells at $z = 0$, and
$2L$ is the distance between the minima of the wells (see Fig.~\ref{fig:dwell}).
We present all of our results in terms of the system's characteristic length
$a_{ho} = (\hbar/ m\omega_{ho})^{1/2}$ and energy $\hbar\omega_{ho}$.

For $\omega_{ho}$ and $\alpha$ fixed, the parameter $L$ can be used to scale the height
of the potential barrier between the wells, $V_{ext}(0)$, which is given by
$\frac{1}{2}m\omega_{ho}^{2}\alpha L^{4}$.
To analyze the behavior of squeezing and fragmentation over a wide range of potentials, we fix
$\alpha = 4/81\, a_{ho}^{-2}$ and choose three different values of $L$,  $L=a_{ho}$, $2\, a_{ho}$,
and $3\, a_{ho}$, giving barrier heights $2/81\,\hbar\omega_{ho}$, $32/81\,\hbar\omega_{ho}$,
and $2\,\hbar\omega_{ho}$, respectively, that range from very small to very high, as seen in
Fig.~\ref{fig:dwell}. Table \ref{tab:zEnergies} lists the energies of the first four states
of the double well for each of these three potentials, to give a sense of where they lie relative
to the height of the barrier. These states will be relevant to constructing models
for the double well system in Sec.~\ref{sec:models}.

\begin{table}
\begin{tabular}{cccccc}
\toprule
& \multicolumn{5}{c}{Energy/$\hbar\omega_{ho}$} \\
\cline{2-6}
$L/a_{ho}$ & $\phi_{0}(z)$ & $\phi_{1}(z)$ & $\phi_{2}(z)$ &  $\phi_{3}(z)$ & $V_{ext}(0)$\\
\colrule
1 & 0.167 & 0.594 & 1.220 & 1.946 & 0.025\\
2 & 0.297 & 0.482 & 1.026 & 1.614 & 0.395\\
3 & 0.634 & 0.637 & 1.681 & 1.801 & 2\\
\botrule
\end{tabular}
\caption{Energies of the ground and first three excited states of the $z$ component
of the double well potential, as well as the energy of the double well barrier, for
$\alpha=4/81\, a_{ho}^{-2}$ and three different potentials parameterized by $L$
(see Sec.~\ref{sec:hamiltonian} and Fig.~\ref{fig:dwell}). Note that these energies do not include
the contribution from the $x$ and $y$ components of the state.}
\label{tab:zEnergies}
\end{table}

We use the experiment described in~\cite{Shin2004} to give a realistic sense of the magnitude
of the parameters of the external potential. This experiment used \cf{^{23}Na} atoms and a trap with
$L=6.5\mathrm{\, \mu m}$, $\omega_{ho}/2\pi = 615\mathrm{\, Hz}$, and
$\frac{1}{2}m\omega_{ho}^{2}\alpha L^{4} = h \times 4.7 \mathrm{\, kHz}$
(equivalently, $\alpha = 6.1\times10^{-9}\mathrm{\, nm}^{-2}$).  Thus, $a_{ho} = 845\mathrm{\, nm}$
and $\hbar\omega_{ho} = 2.54\times10^{-12}\mathrm{\, eV}$, so $L = 7.7 \, a_{ho}$ and
$\alpha = 0.26 \, a_{ho}^{-2}$. These values are comparable to those in the systems we simulate.

We restrict our attention to BECs with repulsive interactions.
Because we are interested in ground state (i.e., low energy) properties, we assume \emph{s}-wave
scattering and hence use a hard sphere interaction potential:
\begin{equation}
V_{int}(r_{jk}) = \left\{
   \begin{array}{ll}
   \infty & r_{jk} \leq a \\
   0 & r_{jk} > a
   \end{array}
\right. ,
\label{eq:hard_sphere_pot}
\end{equation}
where $r_{jk} = | \mathbf{r}_{j}-\mathbf{r}_{k} |$ and $a$ is the positive
\emph{s}-wave scattering length, which determines the effective interaction strength.
Substantial tunability of $a$ has been demonstrated in the laboratory using Feshbach resonances.
A particularly impressive example is~\cite{Pollack2009}, in which $a$ for \cf{^{7}Li} was tuned
between $0.53 \mathrm{\, pm}$ and $10.6 \mathrm{\, \mu m}$
(between $3.5\times 10^{-7} \, a_{ho}$ and $7 \, a_{ho}$, given the value of $\omega_{ho}$ above).
Our calculations employ values of $a$ up to $0.5\,a_{ho}$ (about $420 \mathrm{\, nm}$),
and are thus well within the range of experimental accessibility.

\subsection{The Differential Number Distribution and Squeezing}
\label{sec:numdist}

We are primarily interested in understanding the relationship between
number squeezing and interaction strength.  The differential number distribution
is related to the operator $\hat{n} = \frac{1}{2}(\hat{L}-\hat{R})$,
where $\hat{L}$ is given in the position representation by
\begin{equation}
\hat{L} = \sum_{i=1}^{N}\left\{
   \begin{array}{ll}
   1 & z_{i} < 0 \\
   0 & z_{i} > 0
   \end{array}
\right. ,
\label{eq:LandR}
\end{equation}
and $\hat{R}$ is given by the analogous expression.
For a completely symmetric state $\Psi(\mathbf{r}_{1},\ldots,\mathbf{r}_{N})$, we find
\begin{equation}
\braket{\Psi | \hat{L} | \Psi} = N \int_{-\infty}^{0} \left[\int_{-\infty}^{\infty} \int_{-\infty}^{\infty}| \Psi_{1}(\mathbf{r}_{1})|^{2} dx_{1} dy_{1} \right] dz_{1},
\end{equation}
where
\begin{equation}
|\Psi_{1}(\mathbf{r}_{1})|^{2} = \int_{-\infty}^{\infty} \cdots \int_{-\infty}^{\infty}|\Psi(\mathbf{r}_{1},\ldots,\mathbf{r}_{N})|^{2} d\mathbf{r}_{2} \cdots d\mathbf{r}_{N},
\end{equation}
and similarly for $\braket{\Psi | \hat{R} | \Psi}$. Intuitively,
$\hat{L}$ and $\hat{R}$ measure the fraction of the
probability density of the many-body state that exists in the left and right wells, respectively.

Because of the symmetry of the double well potential, the ground state of the system has
$\braket{\Psi | \hat{L} | \Psi} = \braket{\Psi | \hat{R} | \Psi} = N/2$
(and hence $\braket{\Psi | \hat{n} | \Psi} = 0$) regardless of the strength
of the interaction between the particles.  However, the width of this distribution,
characterized by its standard deviation
$\sigma_{n} = \sqrt{\braket{\Psi | \hat{n}^{2} | \Psi}}$, does vary with $a$.
For the noninteracting ($a=0$) case, the many-body ground state consists
of a product of one-body ground states, and the particles are distributed
according to  a binomial distribution, with $\sigma_{n} = \sqrt{N}/2$.
For a repulsive interaction ($a>0$), we expect that the number distribution will narrow
because configurations with many particles on one side of the double well
and few on the other will be energetically disfavored relative to configurations that more
evenly split the particles between the two sides.
This narrowing is what is meant by number squeezing. We define a squeezing parameter $S$
to characterize the amount of squeezing in the system relative to the non-interacting ground state:
\begin{align}
S &= 1-\frac{\sigma_{n}^{2}}{N/4} \nonumber\\
&= 1 - \frac{1}{N} \braket{\Psi | (\hat{L} - \hat{R})^{2} | \Psi}.
\label{eq:S}
\end{align}
$S=0$ corresponds to no squeezing, and $S=1$ corresponds to a fully squeezed state, in which $\sigma_{n}=0$.

Based on this qualitative argument, we would expect squeezing to increase with
interaction strength.  This is indeed the prediction of the two-mode
mean-field model when the two wells are well-separated~\cite{Javanainen1999},
but as we shall see below, it fails to hold when this condition is not met.
The results of both the more accurate eight-mode model and the exact Quantum Monte Carlo simulations
will also be shown to disagree with this simple picture.

\subsection{The One Body Density Matrix, Fragmentation, and Depletion}
\label{subsec:OBDMfrag}

We analyze additional condensate properties by computing and diagonalizing
the one body density matrix (OBDM), which constitutes a valid description
of the BEC at all densities and interaction strengths~\cite{Penrose1956,Leggett2001}.
This computation results in the fraction of particles that are in the BEC and
the state(s) they occupy, i.e., the extent of depletion and fragmentation,
as a function of interaction strength, and provides another way to understand the breakdown
of the mean-field models.

In a non-interacting system, the condensate is defined in terms of a one-body
ground state wavefunction, and the condensate fraction is the ratio of the number of particles
occupying that state to the total number of particles. For a uniform system, momentum is
a good quantum number, and the condensate is associated with the zero momentum state
(this is true even when interactions are introduced, and the full many-body ground state may
no longer be described by a one-body wavefunction). In a finite, non-uniform,
interacting system, neither of these prescriptions apply. Instead, analysis of the OBDM
gives the condensate fraction and corresponding state in terms of the largest eigenvalue
of the OBDM and its corresponding eigenvector~\cite{Penrose1956,Lowdin1955,Lee1957}.

The OBDM, which characterizes the correlations between the particle density at points
$\mathbf{r}$ and $\mathbf{r^{\prime}}$ in a many-body quantum state, is given by~\cite{Lowdin1955}
\begin{equation}
\label{eq:OBDM}
\rho(\mathbf{r},\mathbf{r}^{\prime}) = \braket{\hat{\Psi}^{\dagger}(\mathbf{r}) \hat{\Psi}(\mathbf{r}^{\prime})},
\end{equation}
where $\hat{\Psi}(\mathbf{r})$ is the field operator that annihilates a single particle
at the point $\mathbf{r}$.  $\hat{\Psi}(\mathbf{r})$ can be expanded
in terms of a set of one-body wavefunctions $\phi_{i}(\mathbf{r})$
(the so called ``natural orbitals'') and the corresponding annihilation
operators $\hat{a}_{i}$:
\begin{equation}
\hat{\Psi}(\mathbf{r}) =\sum_{i}\phi_{i}(\mathbf{r})\hat{a}_{i}.
\end{equation}
At $T=0$,  $\rho(\mathbf{r},\mathbf{r}^{\prime})$ is evaluated with respect to the $N$-particle
ground state wavefunction $\Psi_{0}(\mathbf{r}_{1}, \ldots, \mathbf{r}_{N})$, yielding
\begin{align}
\label{eq:1-RDM_defining_relation}
\rho(\mathbf{r},\mathbf{r}^{\prime}) & = \braket{\Psi_{0} | \hat{\Psi}^{\dagger}(\mathbf{r}) \hat{\Psi}(\mathbf{r}^{\prime}) | \Psi_{0}} \nonumber\\
& = \sum_{ij} \phi_{i}^{\ast}(\mathbf{r}) \phi_{j}(\mathbf{r}^{\prime})  \braket{\Psi_{0} | \hat{a}_{i}^{\dagger} \hat{a}_{j} | \Psi_{0}} \nonumber\\
& = \sum_{i} \phi_{i}^{\ast}(\mathbf{r}) \phi_{i}(\mathbf{r}^{\prime}) N_{i}
\end{align}
where $\sum_{i} N_{i} = N$. The natural orbitals may thus be obtained as the
eigenvectors of the OBDM in the position representation, and the corresponding eigenvalues
$N_i$ give the occupation numbers of these natural orbitals in the many-body
ground state.

As a matter of notation, the natural orbital with highest occupation is given an index of $0$,
the next highest an index of $1$, etc, and we denote the fraction of particles occupying
a given natural orbital by $\mathcal{N}_{i}=N_{i}/N$.  Any natural orbital which is occupied
in the thermodynamic limit (i.e., which has nonzero $\mathcal{N}_{i}$ as $N$ approaches infinity)
can be interpreted as a condensate. For a typical BEC, there is only one such natural orbital.
When there is more than one, the BEC is \emph{fragmented}~\cite{Nozieres1982}.  In either case,
the total small population distributed among the other natural orbitals that vanishes
in the thermodynamic limit is known as the \emph{depletion}~\cite{Bogoliubov1947}.
Hence, fragmentation and depletion can be distinguished in principle because the occupation
of individual depleted orbitals goes to zero in the thermodynamic limit, but the fragmented states
maintain a finite occupation. In practice, however, all of our work is done at finite $N$,
so the distinction between fragmentation and depletion is ambiguous.  Below, we provide definitions
for fragmentation and depletion parameters that are appropriate and useful
for the double well system.

Intuitively, fragmentation in the double well can be related to the fluctuation of particles
across the barrier.  Suppose the barrier is very weak; then the ground state of the system is
essentially the ground state of a single well, and there is no fragmentation.  On the other hand,
if the barrier is very strong, so that the wells can be thought of as isolated, then the particles
in each well form independent condensates and the system is highly fragmented.
Indeed,~\cite{Spekkens1999} predicted that the amount of fragmentation observed in
a double well system would increase with the height of the barrier. Additionally,
stronger interactions lead to reduced fluctuations, which constrains each particle in the system
to occupy only one well.  Thus, for strong repulsive interactions the condensate fragments
into two independent condensates. This implies that fragmentation should also increase with
interaction strength for a fixed barrier.

In analogy to the squeezing parameter $S$, we define fragmentation and depletion parameters,
$F$ and $D$. Reference~\cite{Mueller2006} demonstrates that a condensate with
$G$-fold degeneracy in its ground state can fragment into $G$ parts, assuming low degeneracy
($G \approx 1$).  From the energies listed in Table~\ref{tab:zEnergies}, we see that the
one-body ground state has near-degeneracy (i.e., $G\rightarrow 2$) when $L$ becomes large.
Hence, it is reasonable to assume that, for the double well, at most two natural orbitals
participate in fragmentation, and the rest, if occupied, constitute a very small amount
of depletion (i.e., $\mathcal{N}_{0} + \mathcal{N}_{1} \approx 1$).  This motivates the definition
of fragmentation and depletion parameters ($F$ and $D$) as
\begin{align}
F &= 1- |\mathcal{N}_{0}-\mathcal{N}_{1}|\label{eq:F} \\
D &= 1- (\mathcal{N}_{0}+\mathcal{N}_{1})\label{eq:D}.
\end{align}
With these definitions, a single condensate is represented by $F \approx D \approx 0$
and a doubly fragmented condensate is represented by $F \approx 1$ and $D \approx 0$.
Because the OBDM in the two-mode model is a $2\times 2$ matrix, there are only two natural
orbitals and two occupation numbers for the system within that context.  Hence, the depletion
as defined here is necessarily zero for a two-mode description (see Sec.~\ref{subsub:FandDin2mode})
but can be non-zero for an eight-mode description and in the QMC simulations.

\section{Truncated Basis Models for an Interacting BEC in a Double Well}
\label{sec:models}

Several simplified models have been proposed in attempts to reproduce the behavior of
the interacting double well system while avoiding the difficulty of treating the interaction
exactly. For comparison with the exact Quantum Monte Carlo calculations, we will use two models
that represent the Hamiltonian in a truncated basis of one-body states, specifically the
oft-used two-mode model~\cite{Mahmud2002,Milburn1997,Javanainen1999,Bodet2010,Ananikian2006,Jia2008}
and a recently proposed eight-mode model~\cite{Garcia-March2010}.

The hard sphere interaction potential, Eq.~\eqref{eq:hard_sphere_pot}, imposes the constraint
that the wavefunction between two particles be 0 for $r_{ij} \leq a$.
In the low energy limit and for $r_{ij} \geq a$, the wavefunction generated by the hard sphere
potential is identical  to the one that results from replacing this potential with
a contact potential of the form~\cite{Fermi1936,Breit1947}
\begin{equation}
V_{int}(r_{jk}) = \frac{4\pi\hbar^{2}a}{m} \delta(r_{jk}).
\end{equation}
The Hamiltonian for the system is then
\begin{align}
\hat{H} &= \sum_{i=1}^{N} \left(\frac{\mathbf{p}_{i}^{2}}{2m} + V_{ext}(\mathbf{r}_{i})\right) + \frac{4\pi\hbar^{2}a}{m}\sum_{i<j} \delta(r_{jk})\\
&= \int d\mathbf{r}\, \hat{\Psi}^{\dag}(\mathbf{r}) \left(-\frac{\hbar^{2}}{2m}\nabla^{2} + V_{ext}(\mathbf{r})\right)\hat{\Psi}(\mathbf{r}) \nonumber\\
&\quad + \frac{2\pi\hbar^{2}a}{m} \int d\mathbf{r}\,  \hat{\Psi}^{\dag}(\mathbf{r}) \hat{\Psi}^{\dag}(\mathbf{r}) \hat{\Psi}(\mathbf{r}) \hat{\Psi}(\mathbf{r}),
\label{eq:hamiltonian}
\end{align}
where
\begin{equation}
\hat{\Psi}(\mathbf{r}) = \sum_{i=1}^{\infty} \psi_{i}(\mathbf{r}) \hat{a}_{i},
\end{equation}
resulting in the second-quantized form
\begin{equation}
\hat{H} = \sum_{i,j=1}^{\infty} \hat{a}_{i}^{\dag}\hat{a}_{j} \epsilon_{ij} +  a\!\! \sum_{i,j,k,l=1}^{\infty}  \!\! \hat{a}_{i}^{\dag}\hat{a}_{j}^{\dag}\hat{a}_{k}\hat{a}_{l} \kappa_{ijkl},
\label{eq:hamiltonian2}
\end{equation}
with
\begin{align}
\epsilon_{ij} &= \int d\mathbf{r}\, \psi_{i}^{\ast}(\mathbf{r}) \left(-\frac{\hbar^{2}}{2m}\nabla^{2} + V_{ext}(\mathbf{r})\right)\psi_{j}(\mathbf{r})\label{eq:epsilon}\\
\kappa_{ijkl} &= \frac{2\pi\hbar^{2}}{m} \int d\mathbf{r}\,  \psi_{i}^{\ast}(\mathbf{r}) \psi_{j}^{\ast}(\mathbf{r}) \psi_{k}(\mathbf{r}) \psi_{l}(\mathbf{r})\label{eq:kappa}.
\end{align}
The (one-body) energy of a state $i$ is given by $\epsilon_{ii}$.
For $i \neq j$, the $\epsilon_{ij}$ parameters characterize the tunneling between
states $i$ and $j$.  The $\kappa_{ijkl}$ parameters characterize the strength of
two-body interactions. Once the one-body basis $\psi_{i}(\mathbf{r})$ is specified,
these parameters are then solely a function of the geometry of the potential.

In the following subsections, we describe the models obtained by expanding $\hat{\Psi}(\mathbf{r})$
in a truncated basis of either two or eight non-interacting one-body states,
consistent with  the bases employed in most previous work on finite mode representations.
For the two-mode system, we also analyze an alternative basis set constructed
from solutions to the Gross-Pitaevskii (GP) equation as advocated in~\cite{Ananikian2006}.
As we explain in Section~\ref{subsub:GPdetails}, the systematic
study of the effects of varying $N$, $a$, and the number of modes is more accessible
with a non-interacting, one-body basis. To understand the dependence on these parameters
in detail, we therefore present detailed analysis of calculations made with the usual
non-interacting basis that has been employed in most previous work
in Sections \ref{sec:2moderesults} and \ref{sec:8moderesults}.
We add the more accurate results of two-mode calculations made with a GP basis when comparing
the finite basis calculations with results of the Monte Carlo calculations in
Sec.~\ref{subsub:QMCsqueezingresults}.

\subsection{Two-mode model}
\label{subsec:twomode}

\begin{figure*}
\includegraphics{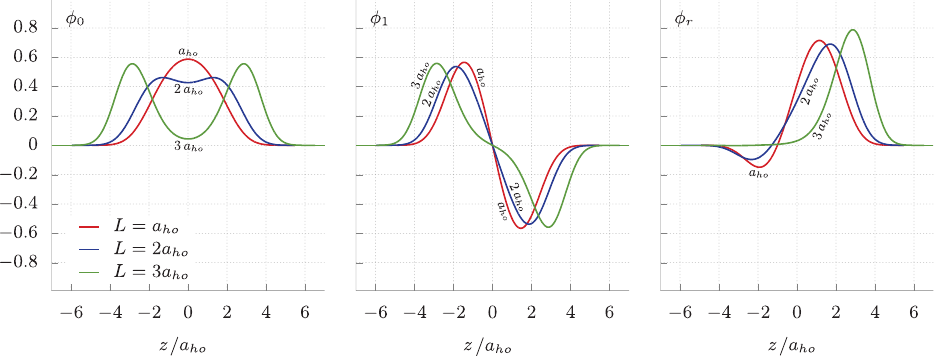}
\caption{\label{fig:psi_2}(Color online) The ground state ($\phi_{0}$, left panel), first excited state ($\phi_{1}$, middle panel), and right localized linear combination of those states ($\phi_{r}$, right panel) for the $z$ component of the double well potential for $\alpha = 4/81\,a_{ho}^{-2}$. Each panel indicates how its respective state varies with $L$, for $L=a_{ho}$, $2\,a_{ho}$, and $3\,a_{ho}$.  We do not plot the left localized linear combination of the ground and first excited states ($\phi_{l}$) because that is simply $\phi_{r}$ reflected about the $z=0$ axis.}
\end{figure*}

The two-mode model includes the lowest two one-body energy states of the 3D double well
\begin{equation}
\label{eq:twomode_modes_ge}
\psi_{g/e}(\mathbf{r}) = \psi_{0}(x,y) \phi_{0/1}(z),
\end{equation}
where $\phi_{0}(z)$ and $\phi_{1}(z)$ are the ground and first excited states,
respectively, of the 1D double well and
\begin{equation}
\psi_{0}(x,y) = \psi_{0}^{ho}(x) \psi_{0}^{ho}(y),
\end{equation}
is the ground state of a 2D harmonic oscillator. As is conventional, we expand
$\hat{\Psi}(\mathbf{r})$ in terms of linear combinations of these states that are
localized in the left and right wells of the potential, i.e.,
\begin{equation}
\label{eq:twomode_modes}
\psi_{l/r}(\mathbf{r}) = \psi_{0}(x,y) \phi_{l/r}(z),
\end{equation}
where
\begin{equation}
\phi_{l/r}(z) = \frac{1}{\sqrt{2}} (\phi_{0}(z) \pm \phi_{1}(z)),
\end{equation}
with corresponding left and right annihilation (creation) operators $a_l$ ($a_l^\dagger$)
and $a_r$ ($a_r^\dagger$), respectively. See Fig.~\ref{fig:psi_2} for examples of $\phi_{0/1}(z)$
and $\phi_{l/r}(z)$ for the three different potentials employed here.

\subsubsection{The two-mode Hamiltonian}
By expanding $\hat{\Psi}(\mathbf{r})$ in terms of Eq.~\eqref{eq:twomode_modes},
the Hamiltonian, Eq.~\eqref{eq:hamiltonian2}, becomes
\begin{align}
\hat{H} &= N (\Delta +  a (N-1)\kappa_{0}) \hat{I} \nonumber\\
&\quad - 2 a(\kappa_{0} - 2\kappa_{2}) \hat{n}_{l}\hat{n}_{r}\nonumber\\
&\quad - (\delta/2 - 2 a(N-1)\kappa_{1})(\hat{a}_{l}^{\dag}\hat{a}_{r}+\hat{a}_{r}^{\dag}\hat{a}_{l}) \nonumber\\
&\quad +  a \kappa_{2} (\hat{a}_{l}^{\dag}\hat{a}_{l}^{\dag}\hat{a}_{r}\hat{a}_{r}+\hat{a}_{r}^{\dag}\hat{a}_{r}^{\dag}\hat{a}_{l}\hat{a}_{l}),
\label{eq:2modeham}
\end{align}
where
$\hat{n}_l \, (\hat{n}_r) = a_l^\dagger a_l \, (a_r^\dagger a_r)$,
$\delta = \epsilon_{ee}-\epsilon_{gg}$ (i.e., the energy difference
between the excited and ground states), $\Delta = (\epsilon_{ee}+\epsilon_{gg})/2$
(i.e., the average energy of the excited and ground states, or equivalently, the
energy of the left and right localized states), and the $\kappa$ parameters
are shorthand for various combinations of the $\kappa_{ijkl}$'s.  In particular,
\begin{align}
\kappa_{0} &= \kappa_{llll} = \kappa_{rrrr} \nonumber\\
&= (\kappa_{gggg} + \kappa_{eeee} + 6\kappa_{ggee})/4 \label{eq:kappa0}\\
\kappa_{1} &= \kappa_{lllr} = \kappa_{lrrr}\nonumber\\
&=(\kappa_{gggg} - \kappa_{eeee})/4 \label{eq:kappa1}\\
\kappa_{2} &= \kappa_{llrr}\nonumber\\
&= (\kappa_{gggg} + \kappa_{eeee} - 2\kappa_{ggee})/4\label{eq:kappa2}.
\end{align}
Since the states are all real, $\phi_{0}(z)$ is even, $\phi_{1}(z)$ is odd, and
$\phi_{l}(z) = \phi_{r}(-z)$, these three parameters constitute the only distinct and
nonzero matrix elements $\kappa_{ijkl}$ in the two-mode model.  Note that both $\kappa_{0}$
and $\kappa_{2}$ are positive, while $\kappa_{1}$ can be positive or negative.  We can estimate
the relative size of these matrix elements by defining the function $\beta(z)$,
\begin{equation}
\beta(z) = \phi_{1}^{2}(z) - \phi_{0}^{2}(z),
\label{eq:beta}
\end{equation}
from which we find
\begin{align}
\kappa_{0} &= 2\int_{-\infty}^{\infty} \phi_{0}^{4}(z) dz + 2\int_{-\infty}^{\infty} \phi_{0}^{2}(z) \beta(z) dz\nonumber\\
& \quad  + \frac{1}{4}\int_{-\infty}^{\infty} \beta^{2}(z) dz \\
\kappa_{1} &= -\frac{1}{2}\int_{-\infty}^{\infty} \phi_{0}^{2}(z) \beta(z) dz - \frac{1}{4}\int_{-\infty}^{\infty} \beta^{2}(z) dz \\
\kappa_{2} &= \frac{1}{4}\int_{-\infty}^{\infty} \beta^{2}(z) dz.
\label{eq:kappas}
\end{align}
When $\phi_{0}^{2}(z) \approx \phi_{1}^{2}(z)$ (i.e., the two modes are nearly degenerate,
as in the case when $L=3\,a_{ho}$), $\beta(z) \ll 1$ and therefore
$\kappa_{0} \gg |\kappa_{1}| \gg \kappa_{2}$.
Otherwise, $\kappa_{0} > |\kappa_{1}| \approx \kappa_{2}$. See Table \ref{tab:kappas}
for representative numerical values.
\begin{table}
\begin{tabular}{ccccc}
\toprule
& & \multicolumn{3}{c}{$\kappa/(\hbar\omega_{ho}/a_{ho}$)} \\
\cline{3-5}
$L/a_{ho}$ & $\delta/\hbar\omega_{ho}$ & $\kappa_{0}$ & $\kappa_{1}$ & $\kappa_{2}$ \\
\colrule
1 & $4.28\times 10^{-1}$ & $5.74\times 10^{-2}$ & $\,\,\,\,9.09\times 10^{-4}$ & $6.79\times 10^{-3}$  \\
2 & $1.85\times 10^{-1}$ & $5.31\times 10^{-2}$ & $-1.22\times 10^{-3}$ & $2.85\times 10^{-3}$  \\
3 & $3.02\times 10^{-3}$ & $6.93\times 10^{-2}$ & $-1.25\times 10^{-4}$ & $1.35\times 10^{-6}$ \\
\botrule
\end{tabular}
\caption{\label{tab:kappas} Values of the energy splitting $\delta$ and two-body interaction
parameters $\kappa_{0}$, $\kappa_{1}$, and $\kappa_{2}$ for three trap geometries
($L=a_{ho}$, $2\,a_{ho}$, and $3\,a_{ho}$).}
\end{table}

The natural basis for the two-mode Hamiltonian is a Fock basis
$\ket{n}$, where $\ket{n}$ consists of the fully-symmetrized state with $n$ particles
in the $\psi_{l}(\mathbf{r})$ state and $N-n$ particles in the $\psi_{r}(\mathbf{r})$
state, i.e., $\ket{n} = \ket{n}_{l}\ket{N-n}_{r}$. We can interpret the terms of the
two-mode Hamiltonian, Eq.~\eqref{eq:2modeham}, in the context of this Fock basis as follows
(where we have omitted the coefficients of the operators for brevity):

\subparagraph{$\hat{I}$:}
The energy that each Fock state has in common.  We will ignore these terms
when analyzing squeezing, because doing so does not alter the ground state
wavefunction of the system.
\subparagraph{$-\hat{n}_{l}\hat{n}_{r}$:}
The energy of each Fock state due to interactions between fixed numbers of particles in each well.
This energy is lower the more evenly distributed the particles are, so $\ket{N/2}$
is the ground state for this term when considered alone.
\subparagraph{$-(\hat{a}_{l}^{\dag}\hat{a}_{r}+ \hat{a}_{r}^{\dag}\hat{a}_{l})$:}
The energy due to transitions between
Fock states that involve a single particle switching from the left to the right mode, or vice versa.
This term depends on the scattering length $a$, in addition to the usual dependence
on the energy gap $\delta$ between the one-body ground
and excited states. The ground state for this part of the Hamiltonian alone is
$\frac{1}{2^{N/2}} \sum_{n=0}^{N} \sqrt{\binom{N}{n}} \ket{n}$. Note that removing the minus
sign in front of this term would leave the magnitude of the coefficients of this state
unchanged but would cause their signs to alternate.
\subparagraph{$\hat{a}_{l}^{\dag}\hat{a}_{l}^{\dag}\hat{a}_{r}\hat{a}_{r}+ \hat{a}_{r}^{\dag}\hat{a}_{r}^{\dag}\hat{a}_{l}\hat{a}_{l}$:}
The energy due to transitions between Fock states that involve exactly two particles switching
from the left to the right mode, or vice versa. This
coherent pair exchange term is due entirely to inter-particle
interactions. A Hamiltonian which includes only these terms can be rewritten in a block-diagonal
form with two blocks, where each block involves either the even-numbered or the odd-numbered
Fock states.  Hence, the ground state can only involve either even or odd Fock states, but not both.
We can confirm this reasoning through an explicit computation of the ground state of this term
alone, which shows it to be
$\frac{1}{2^{N/2}} \sum_{\substack{n=0\\n \in \mathrm{even}}}^{N} \frac{\sqrt{n!(N-n)!}}{(n/2)!((N-n)/2)!} \ket{n}$.

Finally, it is useful to note that within the two-mode Hamiltonian, the intrinsic interaction
strength, given by the $s$-wave scattering length $a$, is scaled by $(N-1)\kappa_0$,
$\kappa_0 - 2\kappa_2$, $(N-1)\kappa_{1}$, and $\kappa_2$, depending which term of the Hamiltonian
is considered. This suggests that the products of these coefficients with $a$ should be treated
as effective interaction strengths for the two-mode model.  Since the parameters $\kappa_i$
implicitly depend on the barrier height via their dependence on integrals over the one-body
ground and first excited state wave functions, these effective interaction strengths will depend
on the barrier height parameter $L$.  Some of them also manifestly depend on $N$,
and all of them depend on these three parameters ($L$, $a$, $N$) in different ways.
Thus, to obtain a full understanding of the independent roles of the barrier height,
intrinsic interaction strength, and number of particles on squeezing and fragmentation,
we shall explicitly study the different and distinct dependencies of number squeezing and fragmentation on $L$, $a$ and $N$ in Section~\ref{sec:results}.

\subsubsection{Nearly degenerate two-mode model}
\label{sec:degenerate2mode}

Several previous studies~\cite{Mahmud2002,Milburn1997,Javanainen1999,Bodet2010} have
analyzed the double well system under the two-mode model with the
assumption that the two modes are nearly degenerate,
so that $\phi_{0}^{2}(z) \approx \phi_{1}^{2}(z)$. Physically, this
can be achieved by imposing a high and/or wide barrier, i.e., large $L$.
Mathematically, this amounts to assuming that $\kappa_{gggg} = \kappa_{ggee} = \kappa_{eeee}$
(whence $\kappa_{1} = \kappa_{2} = 0$), thereby reducing the two-mode Hamiltonian to
\begin{equation}
\hat{H} = -\frac{\delta}{2}(\hat{a}_{l}^{\dag}\hat{a}_{r}+\hat{a}_{r}^{\dag}\hat{a}_{l}) -2 a \kappa_{0} \hat{n}_{l}\hat{n}_{r}\label{eq:JandIHam}.
\end{equation}
When the one-body states lie below the barrier $V_{ext}(0)$ (c.f.~Table~\ref{tab:zEnergies}),
the first term describes one-body tunneling between the left and right potential wells,
quantified by the pure potential parameter $\delta$ (note that the exact two-mode Hamiltonian
of Eq.~\eqref{eq:2modeham} has an additional $a$-, $N$-, and $\kappa_1$-dependent contribution
to this amplitude).

This Hamiltonian has two natural limits.  When tunneling dominates ($a=0$), the ground state
is $\frac{1}{2^{N/2}} \sum_{n=0}^{N} \sqrt{\binom{N}{n}} \ket{n}$ and there is no squeezing ($S=0$).
When interactions dominate ($\delta$ = 0), the ground state is $\ket{N/2}$ and squeezing is maximal
($S=1$). This behavior matches the qualitative argument made in Sec.~\ref{sec:numdist}.

Ref.~\cite{Javanainen1999} used the nearly degenerate
two-mode model to compute an approximate analytical expression for the relative
squeezing $S$. In the notation of the present work, this is given by
\begin{equation}
\label{eq:JandIresult}
S_{nd2} = 1 - 2^{1/3}\left\{
\begin{array}{ll}
\left(\frac{1}{2^{2/3} + N a/a^{\ast}}\right)^{1/2} & a \le a^{\ast}N\\
N \left(\frac{a^{\ast}}{a}\right)^{2} & a > a^{\ast}N
\end{array}
\right. ,
\end{equation}
where $a^{\ast} = \delta/2^{10/3}\pi \kappa_{0}$ is a function of the
geometry of the double well.  As $a$ ranges between zero and infinity,
Eq.~\eqref{eq:JandIresult} predicts that $S_{nd2}$ will vary monotonically between 0 and 1
(apart from a discontinuity of $\mathcal{O}(N^{-3})$ at $a = a^{\ast} N$ that is an
result of the approximations employed in the derivation of $S_{nd2}$~\cite{Javanainen1999}).

\subsubsection{Exact two-mode model}
\label{subsubsec:exacttwomode}

The nearly degenerate two-mode model, while analytically tractable, misses many interesting
features of the double well system that are also necessary to include for an informed comparison
to the exact Quantum Monte Carlo calculations. To identify these features, we therefore analyze
Eq.~\eqref{eq:2modeham} without making the assumption of near-degeneracy between the one-body
ground and first excited states.  We note that the full two-mode Hamiltonian was studied
in \cite{Ananikian2006,Bader2009} and the two-mode Hamiltonian with $\kappa_{2}=0$
in~\cite{Jia2008}, but none of these previous studies included an analysis of squeezing,
which is one of our primary goals.

The ground state of Eq.~\eqref{eq:2modeham} is obtained via numerical diagonalization
in the Fock representation using a restricted basis. Since the size of the Hilbert space
is $N+1$, this diagonalization is tractable for $N$ up to several thousand.
Given the coefficients $c_{n}$ from the expansion of the ground state
(i.e., $\ket{\psi_{gs}} = \sum_{n=0}^{N}c_{n}\ket{n}$), the analytical expression for the
two-mode squeezing $S_{2}$ is
\begin{align}
S_{2} &= S_{max}\times\left(1 - \frac{1}{N} \braket{\psi_{gs} | (\hat{n}_{l}-\hat{n}_{r})^{2} | \psi_{gs}}\right)\nonumber\\
&= S_{max}\times\left(1 - \frac{1}{N} \sum_{n=0}^{N}(N-2n)^{2}|c_{n}|^{2}\right),
\label{eq:squeezing2}
\end{align}
with
\begin{equation}
S_{max} = \left(1- 2\int_{-\infty}^{0} |\phi_{l}(z)|^{2} \,dz \right)^{2}.
\label{eq:Smax}
\end{equation}
$S_{max}$ is the largest value of $S_{2}$ achievable for a given potential and is achieved only
in the pure $\ket{N/2}$ state. For $\alpha=4/81\,a_{ho}^{-2}$, explicit evaluation of this
expression yields values of $S_{max} = 0.704$, 0.820, and 0.999 for $L=a_{ho}$, $2\,a_{ho}$,
and $3\,a_{ho}$, respectively; as the barrier is increased, $S_{max}$ increases to approach
1.  The extent of squeezing in the exact two-mode model is therefore controlled by both
the value of $S_{max}$, which is determined solely by the well geometry, and the composition
of the ground state (i.e., the $c_{n}$ parameters), which is determined by both the well geometry
and the interaction strength $a$.

For finite barriers, $S_{max}$ is quite sensitive to the barrier height because it is a function
of the degree of degeneracy of the two modes, which is controlled by the well geometry.
When the modes are exactly degenerate, then all of the probability for $\phi_{l}(z)$ is contained
in the left well and $S_{max}=1$. Otherwise, part of $\phi_{l}(z)$ extends into the right side
of the potential (and vice versa for $\phi_{r}(z)$), and $S_{max} < 1$. Physically, this means that
when the modes are not exactly degenerate, there is a nonzero probability of measuring a difference
in the number of particles between the two wells, even in the $\ket{N/2}$ state.  Mathematically,
this means that $\hat{n}_{l} = \hat{L}$ and $\hat{n}_{r} = \hat{R}$ only at complete degeneracy,
so that $S_{2}$ is then equivalent to $S$ as defined in Eq.~\eqref{eq:S}. One can thus interpret
$S_{max}$ as compensating for the fact that in the non-degenerate case (i.e., for realistic finite
barrier heights), the operators in the two-mode Hamiltonian ($\hat{n}_{l}$ and $\hat{n}_{r}$)
are not identical to the operators that define squeezing ($\hat{L}$ and $\hat{R}$,
see Eq.~\eqref{eq:LandR}).

An additional difference between the degenerate and non-degenerate cases is the dependence of the
one- and two-body tunneling amplitudes on $a$ in Eq.~\eqref{eq:2modeham} as compared
with Eq.~\eqref{eq:JandIHam}.  In the degenerate case, $\ket{N/2}$ is the ground state in the
$a \rightarrow \infty$ limit, but this is no longer true in the non-degenerate case.  Indeed,
the degeneracy of the system influences the composition of the ground state (and thus the amount
of squeezing) for all nonzero values of $a$.  The results presented in Sec.~\ref{sec:2moderesults}
will show that for the exact two-mode model, not only is $\ket{N/2}$ not the ground state in the
$a \rightarrow \infty$ limit, but also that squeezing does not necessarily increase monotonically
as a function of $a$.

\subsubsection{Fragmentation and depletion in the two-mode model}
\label{subsub:FandDin2mode}

We can also study fragmentation in the two-mode model by expanding the OBDM, Eq.~\eqref{eq:OBDM},
in terms of the left/right localized wave functions, Eqs.~\eqref{eq:twomode_modes},
and diagonalizing to find the occupation of the natural orbitals.  This results in
\begin{equation}
\rho(\mathbf{r},\mathbf{r}^{\prime}) = \mathcal{N}_{0} \phi_{0}(\mathbf{r}) \phi_{0}(\mathbf{r}^{\prime}) + \mathcal{N}_{1} \phi_{1}(\mathbf{r}) \phi_{1}(\mathbf{r}^{\prime}),
\end{equation}
where the fractional occupations are given by
\begin{equation}
\label{eq:2mode_occupations}
\mathcal{N}_{0/1} = \frac{1}{2} \pm \frac{1}{N}\sum_{n=0}^{N-1}\sqrt{(N-n)(n+1)}\, c_{n}c_{n+1}.
\end{equation}
Thus, in the two-mode model, the natural orbitals are the one-body ground and
first excited states of the double well, and their occupations depend on the composition
of the ground state of the system. The fragmentation and depletion parameters,
Eqs.~\eqref{eq:F} and \eqref{eq:D}, are then given by
\begin{subequations}
\begin{align}
F &= 1 - \frac{2}{N}\left|\sum_{n=0}^{N-1}\sqrt{(N-n)(n+1)}\, c_{n}c_{n+1}\right|\label{eq:F2} \\
D &= 0.\label{eq:D2}
\end{align}
\end{subequations}
The zero value of $D$ is consistent with its definition (see Section~\ref{subsec:OBDMfrag}).
For the maximally-squeezed state $\ket{N/2}$, $F = 1$ and the system is
maximally fragmented.  For the noninteracting ground state,
the sum evaluates to $N/2$, so that $F = 0$ and there is no fragmentation.
We note that there is no general one-to-one relationship
between the squeezing parameter $S$ and the fragmentation parameter $F$ in the two-mode model.

\subsubsection{Using a mean field basis for the two-mode model}
\label{subsub:GPdetails}

In addition to non-interacting basis states, we also studied the two-mode model using a mean field
basis that consisted of solutions to the Gross-Pitaevski equation.  We used GPELab, a MATLAB toolbox
for computing the ground state and dynamics of Gross-Pitaevskii equations~\cite{Antoine2014},
to evaluate both the fully symmetric GP ground state $\psi_{0}^{GP}$ and the antisymmetric GP
first excited state $\psi_{1}^{GP}$; we obtained the latter by imposing a node in the wavefunction
along the double well $z$-axis when we ran the GPELab code.  These GP basis states were then
inserted into Eqs.~\eqref{eq:epsilon} and \eqref{eq:kappa} to evaluate the Hamiltonian parameters
for the two-mode model. The ground state properties were then obtained following numerical
diagonalization as in Section~\ref{subsubsec:exacttwomode}.

This procedure yields a pair of basis states (i.e., $\psi_{0/1}^{GP}$) that depends implicitly
on both $N$ and $a$.  One expects that such dependence has the benefit that some average effects
of interactions are already incorporated in the basis functions, which may lead to improvement in
two-mode results relative to those obtained from calculations with non-interacting basis states.
However, even for the two-mode representation, a basis of GP states has the detrimental feature
of adding complexity to the calculations, since the basis states and Hamiltonian parameters
have to be recomputed for each value of $N$ and $a$.  Furthermore, for a fragmented system,
one would ideally wish to make a self-consistent calculation of the GP solutions $\psi_{0/1}^{GP}$,
with their respective occupancies $\mathcal{N}_{0/1}$. This can be done with, e.g.,
the technique developed in Ref.~\cite{Sinatra2000}, but it becomes computationally challenging
for a number of the parameter regimes studied here.  Therefore, we restrict ourselves to the
simplest approach in which both ground and excited mean field states are computed independently with $N$ particles.

\subsection{Eight-mode model}
\label{subsec:8mode}
\begin{figure*}
\includegraphics{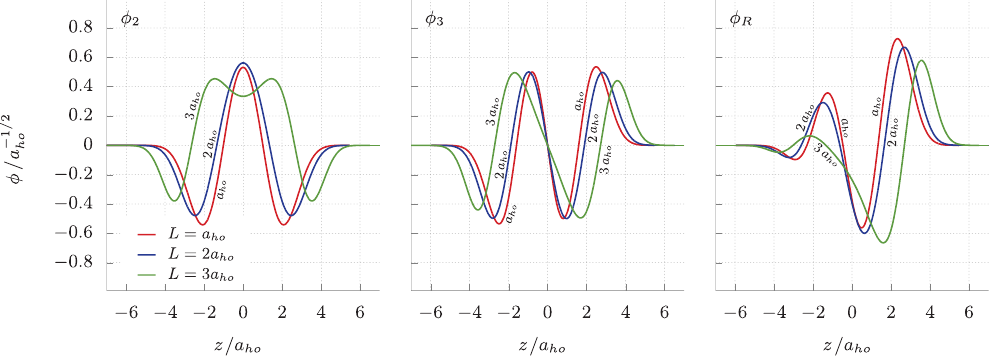}
\caption{\label{fig:psi_8}(Color online) The second excited state ($\phi_{2}$, left panel), third excited state ($\phi_{3}$, middle panel), and right localized linear combination of those states ($\phi_{R}$, right panel) for the $z$ component of the double well potential for $\alpha = 4/81\,a_{ho}^{-2}$. Each panel indicates how its respective state varies with $L$, for $L=a_{ho}$, $2\,a_{ho}$, and $3\,a_{ho}$.  We do not plot the left localized linear combination of the second and third excited states ($\phi_{L}$) because that is simply $\phi_{R}$ reflected about the $z=0$ axis.}
\end{figure*}

In an attempt to improve on the two-mode model, a recently-proposed
eight-mode model~\cite{Garcia-March2010} expands $\hat{\Psi}(\mathbf{r})$ in terms of
the two usual modes, Eqs.~\eqref{eq:twomode_modes}, and six additional modes.
These additional modes may be constructed using left- and right-localized
linear combinations of the second and third double well excited states,
$\phi_{2}(z)$ and $\phi_{3}(z)$:
\begin{equation}
\phi_{L/R}(z) = \frac{1}{\sqrt{2}} (\phi_{2}(z) \pm \phi_{3}(z)).
\end{equation}
See Fig.~\ref{fig:psi_8} for examples of these states.  Constructing additional
two-dimensional harmonic oscillator states in the $xy$-plane,
\begin{equation}
\psi_{\pm}(x,y) = \frac{1}{\sqrt{2}} (\psi_{0}^{ho}(x) \psi_{1}^{ho}(y) \pm i \psi_{1}^{ho}(x) \psi_{0}^{ho}(y)),
\end{equation}
allows the eight modes to be written as
\begin{align}
\label{eq:eightmode_modes}
\psi^{l/r}_{100}(\mathbf{r}) &= \psi_{0}(x,y) \phi_{l/r}(z) \nonumber\\
\psi^{l/r}_{210}(\mathbf{r}) &= \psi_{0}(x,y) \phi_{L/R}(z) \nonumber\\
\psi^{l/r}_{21\pm1}(\mathbf{r}) &= \psi_{\pm}(x,y) \phi_{l/r}(z),
\end{align}
where we have introduced a set of three subscripts to distinguish the modes.
For a potential such that the curvature at the
well minima is roughly equal in the $x$, $y$, and $z$ directions (as opposed
to a pancake- or cigar-like geometry), the subscripts on $\psi_{nlm}^{l/r}$
can be interpreted as the quantum numbers for a particle in a spherical
potential~\cite{Friedman1971}: $n$ indexes the energy of the state and $l$ and $m$ its
angular momentum magnitude and $z$-projection.

We note that since the relative energies of the one-body states depend on the detailed
three-dimensional geometry of the double well (i.e., on the parameters $\omega_{ho}$, $\alpha$,
and $L$), then it is important that these parameters be chosen to ensure that the eight states,
Eqs.~\eqref{eq:eightmode_modes}, do have the lowest energies. The geometries used
in the current work satisfy this property at the same time as they illustrate the behavior
of the system over a wide range of barrier strengths. Table \ref{tab:modeEnergies} lists
the energies of the modes for our three double well geometries.
\begin{table}
\begin{tabular}{ccccc}
\toprule
& \multicolumn{4}{c}{Energy/$\hbar\omega_{ho}$} \\
\cline{2-5}
$L/a_{ho}$ & 100 & 21$\pm$1 & 210 &  $V_{ext}(0)$\\
\colrule
1 & 1.38 & 2.38 & 2.58 & 0.025\\
2 & 1.39 & 2.39 & 2.32 & 0.395\\
3 & 1.64 & 2.64 & 2.74 & 2\\
\botrule
\end{tabular}
\caption{Energies of the 8 modes and the double well barrier,
for potentials with $\alpha=4/81\,a_{ho}^{-2}$ and three different $L$ values.}
\label{tab:modeEnergies}
\end{table}

As in the two-mode case, we use these eight modes to express the Hamiltonian
in a Fock basis and diagonalize the resulting matrix, which we do not reproduce here.
For $N$ particles and $M$ modes, the dimension of the Hilbert space is $\frac{(N+M-1)!}{N!(M-1)!}$,
so the size of the matrix grows rapidly with $N$ for $M=8$.  This restricts
the practical usefulness of the eight-mode model to small $N$ because
of the computational cost of diagonalizing large matrices. However, the Hamiltonian
has a block-diagonal form, with each block corresponding to a different value
of $m$ from $-l$ to $l$. The ground state has $m=0$, so one can make the diagonalization
process easier by only diagonalizing the $m=0$ block. In this way, we were
able to compute eight-mode model results for up to $N=10$.
We do not attempt to apply a GP basis to the eight-mode model since this
would introduce significant additional computational challenges in the construction
of an orthogonal set of eight GP states for each value of $N$ and $a$.

While one can in principle compute analytical expressions for the squeezing, fragmentation,
and depletion parameters in the eight-mode model that are analogous to Eqs.~\eqref{eq:squeezing2},
\eqref{eq:F2}, and \eqref{eq:D2},  the resulting expressions are not particularly illuminating.
Instead, we analyze the differences in these properties between the two models
with numerical calculations in Sec.~\ref{sec:results}.

\section{The Path Integral Ground State Method}
\label{sec:PIGS}

We use the path integral ground state (PIGS) Quantum Monte Carlo method to make exact numerical
calculations of the ground state properties of our system that go beyond the constraints of
the two- and eight-mode models.  PIGS is a many-body, ground state ($T=0$) method that
uses imaginary time propagation and path sampling techniques to calculate the exact
ground  state expectation value for observables in a quantum system.

Conceptually, PIGS starts with a trial wave function that may be written as a sum over
the energy eigenstates of the system: $\ket{\psi_{T}} = \sum_{n=0}^{\infty} c_{n} \ket{\psi_{n}}$.
After applying the operator $e^{-\tau\hat{H}}$, normalizing, and taking the
$\tau \rightarrow \infty$ limit,
the trial wave function decays into the ground state wave function:
\begin{equation}
\lim_{\tau\rightarrow\infty} \frac{e^{-\tau\hat{H}} \ket{\psi_{T}}}{\sqrt{\braket{\psi_{T} |e^{-2\tau\hat{H}} | \psi_{T}}}} = \ket{\psi_{0}} .
\end{equation}
We will ignore the normalization factor for the rest of this discussion.

The configuration of the system is denoted by a $3N$-dimensional vector $R$ that encodes
the coordinates of the system's $N$ particles:
$R \equiv \{\mathbf{r}_{1}, \mathbf{r}_{2}, \ldots, \mathbf{r}_{N}\}$.
In the position representation, it is generally not possible to express
$\braket{R | e^{-\tau\hat{H}} | R^{\prime}}$ analytically unless $\tau$ is small.  Therefore,
we define $\tau \equiv \beta M$, with $\beta \ll 1$.  The expectation value for an observable
of interest $\hat{A}$ (assumed to be diagonal in the position basis) can then be written as
\begin{equation}
\label{eq:diagonalobs}
\begin{split}
\braket{\hat{A}} = {} & \braket{\psi_{T} | (e^{-\beta\hat{H}})^{M} \hat{A} (e^{-\beta\hat{H}})^{M} | \psi_{T}} \\
= {} & \int dR_{0} \cdots dR_{2M} A(R_{M}) \psi^{\ast}_{T}(R_{0}) \psi_{T}(R_{2M}) \\
& \times \prod_{i=0}^{2M-1} G(R_{i}, R_{i+1}, \beta),
\end{split}
\end{equation}
where we have inserted $2M+1$ complete sets of position eigenstates.
Here, $G(R_{i}, R_{i+1}, \beta)=\braket{R_{i} | e^{-\beta\hat{H}} | R_{i+1}}$ is the short time
propagator, which may be approximated for sufficiently small $\beta$.  This procedure can
compute the expectation values of observables in the ground state, but not the ground
state wave function itself.

Given an (exact or approximate) analytical form for $G(R_{i}, R_{i+1}, \beta)$,
our problem has been transformed into that of solving an integral of very high dimension,
which can be done with standard Monte Carlo sampling techniques. The paths
$X \equiv \{R_{0}, R_{1}, \ldots, R_{2M} \}$ are statistically sampled from the probability density
\begin{equation}
\label{eq:probdensity}
\pi(X) = \psi^{\ast}_{T}(R_{0}) \psi_{T}(R_{2M}) \prod_{i=0}^{2M-1} G(R_{i}, R_{i+1}, \beta),
\end{equation}
using the Metropolis algorithm~\cite{Metropolis1953}, which ensures that the sampling is ergodic
(i.e., that the set of accepted paths is a representative sample of the set of all paths).
If this condition is met, then the average value
of $A(R_{M})$ for the set of accepted paths can be used to estimate the value of $\braket{\hat{A}}$.
In general, long paths (large $M$) are required to ensure that $R_{M}$
is sampled from a probability density as close to the square of the exact
ground state wave function as possible.

The main computational difficulty faced when simulating interacting
bosons in double well potentials is  properly estimating the squeezing $S$
for high barriers, because in this situation it is extremely difficult to achieve
ergodicity with respect to moving particles across the barrier. Below
we describe computational details regarding the trial function, the propagator,
methods for path sampling, and the computation of off-diagonal observables,
with explicit consideration of this issue.

\subsection{Trial function}
\label{sec:trialfunction}

We use a trial wave function which is a product of one-body ground state
wave functions and pair correlation (Jastrow) terms:
\begin{equation}
\psi_{T}(R) = \prod_{i=1}^{N}\psi_{0}(\mathbf{r}_{i}) \prod_{j<k}^{N} \left(1-\frac{a}{r_{jk}}\right),
\end{equation}
where $\psi_{0}(\mathbf{r}_{i}) = \psi_{0}^{HO}(x_{i})\psi_{0}^{HO}(y_{i})\psi_{0}^{DW}(z_{i})$,
a product of the analytical harmonic oscillator ground state wave function
in the $x$ and $y$ directions and a numerically calculated one-dimensional double well
ground state wave function in the $z$ direction.  The pair correlation term is the
exact zero-energy \emph{s}-wave scattering solution for two hard spheres~\cite{Lee1957}.

\subsection{Propagator}
\label{sec:propagator}

For our short-time propagator, we use a hybrid form that combines a fourth-order propagator
decomposition with a modification of the free particle propagator that exactly incorporates
the hard sphere interaction.

\subsubsection{External potential decomposition}
First, we use a fourth-order factorization to approximate $G(R_{i}, R_{i+1}, \beta)$~\cite{Chin2002}:
\begin{equation}
\label{eq:G_4th}
\begin{split}
G(R_{i}, R_{i+1}, \beta)  = {} & \int dR_{j} \hspace{2mm} e^{-\frac{\beta}{6}V(R_{i})} \braket{R_{i} | e^{-\frac{\beta}{2}(T+V_{hs})} | R_{j}}\\
& \times e^{-\frac{2\beta}{3} \tilde{V}(R_{j})} \braket{R_{j} | e^{-\frac{\beta}{2}(T+V_{hs})} | R_{i+1}} \\
& \times e^{-\frac{\beta}{6} V(R_{i+1})},
\end{split}
\end{equation}
where $T$ is the kinetic energy, $V$ is the external potential, $V_{hs}$ is the
hard-sphere potential,
\begin{align}
 \tilde{V} &= V + \frac{\tau^{2}}{48} [V, [(T+V_{hs}),V]] \nonumber\\
 &= V + \frac{\lambda\tau^{2}}{24} |\nabla V|^{2},
 \label{eq:Vtilde}
 \end{align}
and $\lambda = \frac{\hbar^{2}}{2m}$.  It is essential to group $V_{hs}$ with $T$ rather than $V$
in the computation of $\tilde{V}$, in order to take advantage of the fact that $[V,V_{hs}]=0$ and
thus avoiding the gradient of the (singular) hard sphere potential.

With this factorization we have introduced a new configuration
$R_{j}$ between each pair of original configurations $R_{i}$ and $R_{i+1}$, so that there are now
$4M + 1$ configurations instead of $2M + 1$.  Treating all of these on equal footing, we can
rewrite Eq.~\eqref{eq:probdensity} as
\begin{equation}
\begin{split}
\pi(X) = {} & \psi^{\ast}_{T}(R_{0}) \psi_{T}(R_{4M}) e^{\frac{1}{6}\beta(V(R_{0}) - V(R_{4M}))}\\
& \times  \prod_{i=0}^{4M-1}  f(R_{i}) \, G_{hs}(R_{i},R_{i+1},\beta/2),
\end{split}
\end{equation}
where
\begin{equation}
f(R_{i}) = \left\{
\begin{array}{lr}
e^{-\frac{1}{3}\beta V(R_{i})} & i = 0,2,\ldots \\
e^{-\frac{2}{3}\beta V(R_{i}) - \frac{1}{36}\lambda\beta^{3}|\nabla V(R_{i})|^{2}} & i = 1,3,\ldots
\end{array}
\right. ,
\end{equation}
and $G_{hs}(R_{i},R_{i+1},\beta/2) = \braket{R_{i} | e^{-\frac{\beta}{2}(T+V_{hs})} | R_{i+1}}$
is the hard sphere propagator.

\subsubsection{Hard sphere propagator}

To compute the hard sphere propagator, we use the pair product approximation~\cite{Barker1979}:
\begin{equation}
\begin{split}
G^{m}_{hs}(R_{i},R_{i+1},\beta) = {} & G^{m}_{free}(R_{i},R_{i+1},\beta)\\
& \times  \prod_{j<k}^{N} \frac{G^{m/2}_{hs}(r^{i}_{jk},r^{i+1}_{jk},\beta)}{G^{m/2}_{free}(r^{i}_{jk},r^{i+1}_{jk},\beta)}.
\end{split}
\label{eq:G_hs_full}
\end{equation}
Here $G^{m/2}_{free/hs}$ is the free/hard sphere propagator for the relative motion between
two particles (a function of the relative coordinates $r_{ij}$ and the reduced mass $m/2$).

Several methods have been proposed in the literature for approximating $G_{hs}$, including
the image approximation~\cite{Barker1979,Jacucci1983} and the propagator
of Cao and Berne~\cite{Cao1992}.  One critical consideration for choosing a propagator
for the double well system is that we need long paths to ensure that the system has decayed
to the ground state because the decay goes as $\exp(-\tau \Delta E)$, where $\Delta E$
(the energy gap between the ground and first excited state) is small. Hence, we must use as large
a time step as possible. We therefore implement the exact hard sphere propagator here,
because it allows larger time steps than the Cao and Berne propagator
(e.g., $10^{-2} \, (\hbar\omega_{ho})^{-1}$ compared to $10^{-4} \, (\hbar\omega_{ho})^{-1}$
for equivalent results).

The exact expression for $G^{m/2}_{hs}$ is the non-closed form~\cite{Larsen1968}:
\begin{equation}
\begin{split}
G_{hs}^{m/2} = {} & \frac{1}{2\pi^{2}} \sum_{l=0}^{\infty} P_{l}(\cos\gamma)(2l+1)\\
& \times \int_{0}^{\infty} k^{2} e^{-2\beta\lambda k^{2}} \frac{R_{l}(r^{i}_{jk},k)R_{l}(r^{i+1}_{jk},k)}{D_{l}(k)}\, dk,
\end{split}
\label{eq:G_hs}
\end{equation}
where
\begin{align}
R_{l}(r,k) &= j_{l}(kr) y_{l}(ka) - y_{l}(kr) j_{l}(ka),\\
D_{l}(k)&= j_{l}^{2}(ka) + y_{l}^{2}(ka),
\end{align}
$j_{l}(x)$ and $y_{l}(x)$ are spherical Bessel functions, and $\gamma$ is the angle
between $r^{i}_{jk}$ and $r^{i+1}_{jk}$.
To use this expression for $G^{m/2}_{hs}$, we must terminate the sum at some
appropriate $l_{max}$ and tabulate it as a function of $r^{i}_{jk}$, $r^{i+1}_{jk}$, and $\gamma$.

An efficient computational representation of Eq.~\eqref{eq:G_hs_full} may be achieved
by rewriting the conventional closed form of $G_{free}^{m/2}$
\begin{equation}
\label{eq:G_free}
G_{free}^{m/2} = \frac{1}{(8\pi\beta\lambda)^{3/2}} e^{-\frac{(\mathbf{r}_{jk}^{i}-\mathbf{r}_{jk}^{i+1})^{2}}{8\beta\lambda}},
\end{equation}
as a summation similar to Eq.~\eqref{eq:G_hs}, namely,
\begin{equation}
\label{eq:G_free_like_hs}
\begin{split}
G_{free}^{m/2} = {} & \frac{1}{2\pi^{2}} \sum_{l=0}^{\infty} P_{l}(\cos\gamma)(2l+1)\\
& \times \int_{0}^{\infty} k^{2} e^{-2\beta\lambda k^{2}} j_{l}(k r^{i}_{jk}) j_{l}(k r^{i+1}_{jk})\, dk.
\end{split}
\end{equation}
The \emph{difference} between $G_{hs}^{m/2}$ and $G_{free}^{m/2}$ converges with respect
to $l_{max}$ much faster than $G_{hs}^{m/2}$ alone.
\begin{figure}
\includegraphics{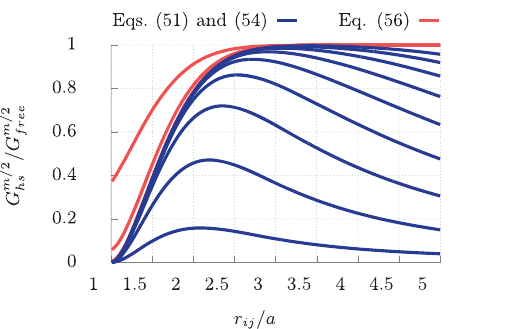}
\caption{(Color online) Comparison of convergence with respect to $l_{max}$ of the ratio
$G^{m/2}_{hs}/G^{m/2}_{free}$ computed using conventional expressions (Eqs.~\eqref{eq:G_hs}
and~\eqref{eq:G_free}, blue curves) versus using the difference expression
(Eq.~\eqref{eq:G_hs_final}, red curves), for $r_{jk}^{i} = r_{jk}^{i+1}$, $\gamma = 0$, and
$a^{2} = 2\beta\lambda$. Both sets of curves are computed for integer values of $l_{max}$ from 0
to 8; the $l_{max} = 0$ curve is the highest (lowest) curve in the red (blue) set. The red curves
with $l_{max}$ between 2 and 8 are visually indistinguishable.  The more rapid convergence
of this ratio when computed using the difference expressions compared with the conventional
expressions is a general feature of these functions. For viewers of the figure in gray scale, the red curves appear as lighter gray and are above the blue curves, which appear as darker gray.}
\label{fig:HSconvergence}
\end{figure}
We can use this fact to reexpress the quotient in the pair product approximation:
\begin{align}
\label{eq:G_hs_final}
\frac{G^{m/2}_{hs}}{G^{m/2}_{free}} = {} & 1 - \frac{G^{m/2}_{free} - G^{m/2}_{hs}}{G^{m/2}_{free}}\nonumber\\
= {} & 1 - \frac{4(2\beta\lambda)^{3/2}}{\pi^{1/2}}  e^{\frac{(\mathbf{r}_{jk}^{i}-\mathbf{r}_{jk}^{i+1})^{2}}{8\beta\lambda}} \nonumber\\
& \times \sum_{l=0}^{\infty} P_{l}(\cos\gamma)(2l+1) \nonumber\\
& \times \int_{0}^{\infty} k^{2} e^{-2\beta\lambda k^{2}} A_{l}(k,r_{jk}^{i},r_{jk}^{i+1}) \, dk,
\end{align}
where
\begin{equation}
A_{l}(k,r,r^{\prime}) = j_{l}(kr)j_{l}(kr^{\prime})- \frac{R_{l}(r,k)R_{l}(r^{\prime},k)}{D_{l}(k)}.
\end{equation}
By using this alternative form for the ratio in Eq.~\eqref{eq:G_hs_full}, we can terminate
the sum at a value of $l_{max}$ about 2 to 10 times smaller (depending on the value of $a$)
than would be necessary to achieve the same precision using Eq.~\eqref{eq:G_hs}.  This is
demonstrated explicitly in Fig.~\ref{fig:HSconvergence}.

\subsection{Sampling methods}
\label{sec:sampling}

For simulations with a double well trapping potential, the key sampling issue is achieving
ergodicity with respect to the motion of particles between the two wells. In general, paths which
are entirely located in one well are more probable than paths that cross the barrier, because of the
extra potential energy associated with the parts of the paths that are in the barrier
region. The larger this difference in probability, the less likely it will be that paths
that start in one well will move to the other over the course of the simulation.
Instead, the paths are often stuck on one side. This problem becomes worse for larger barriers,
and is also exacerbated for small $N$ and small $a$, which one can intuitively understand
as follows. The number of paths is given by $N$, and the minimum distance between different paths
at the same time slice is given by $a$.  When both of these are small, paths can ``settle down''
into the bottoms of the wells where the potential energy is low. However, when either $N$ or $a$
are large, the paths are forced to spread out into regions where the potential is larger,
which makes it easier for them to transition through the barrier because the ``probability penalty''
incurred is not as great.

This ergodicity problem impacts the computation of $S$ more severely than other observables
such as the energy or density. Since $S$ is a function of $(n_{l}-n_{r})^{2}$,
which only changes value when the center of a path crosses the barrier at $z=0$, reduced
ergodicity with respect to particle motion across the barrier leads to long autocorrelation times
for $S$.  Hence, one must wait an unusually long time before the simulation generates enough
independent values of $(n_{l}-n_{r})^{2}$ to compute precise values
of the squeezing $S$.

Here we describe sampling methods that mitigate this problem in certain circumstances.

\subsubsection{Brownian Bridge Moves}

The main ``workhorse'' update method we use is the Brownian bridge move, which is a
specific realization of the more general L\'evy construction~\cite{Levy1940,Ceperley1995}.
In the Brownian bridge move, a portion of the path of a single particle is updated.
The particle is chosen randomly, as is the section of its path that is updated;
the length of this section is a fixed parameter $K$, defined such that the section consists of
$K+1$ time slices, including the endpoints.

The move proceeds as  follows.  The endpoints of the section to be updated are chosen and held
fixed; call these $\mathbf{r}_{0}$ and $\mathbf{r}_{K}$.  Next, the coordinate of the particle
at the first time slice, $\mathbf{r}_{1}$, is replaced with one drawn from the
probability distribution
\begin{equation}
P(\mathbf{r}_{1}) \propto e^{-\frac{\hbar(\mathbf{r}_{1}-\mathbf{r}^{\ast})^{2}}{4\lambda\tau^{\ast}}},
\end{equation}
where
\begin{subequations}
\begin{align}
\mathbf{r}^{\ast} &= \frac{\tau_{1}\mathbf{r}_{0} + \tau_{2}\mathbf{r}_{K}}{\tau_{1} + \tau_{2}}\\
\tau^{\ast} &= \frac{\tau_{1}\tau_{2}}{\tau_{1} + \tau_{2}}.
\end{align}
\end{subequations}
Here, $\tau_{1}$ is the amount of imaginary time separating $\mathbf{r}_{0}$ and $\mathbf{r}_{1}$
and $\tau_{2}$ is the amount of imaginary time separating $\mathbf{r}_{1}$ and $\mathbf{r}_{K}$.
This coordinate becomes the new left endpoint for a section of length $K-1$ that runs from
$\mathbf{r}_{1}$ to $\mathbf{r}_{K}$. The coordinate of the particle at the second time slice,
$\mathbf{r}_{2}$, is replaced with one drawn using a probability distribution with the same form
as the one used for determining $\mathbf{r}_{1}$, but with the updated left endpoint.  This process
continues until the entire section of path is reconstructed.

In general, the Brownian bridge move is an efficient way of sampling new paths, although it is
susceptible to the ergodicity problem described above if the barrier is too strong and enough of
the new path ends up in the barrier region.  For the vast majority of our simulations, however,
it was the only update method that was necessary.

\subsubsection{Swap moves}

One potential way to address the ergodicity issue is to implement an additional type of move
that explicitly transfers a particle from one well to the other~\cite{Krauth2006}.
In our implementation of this ``swap move,'' the $z$-coordinate is negated for the
entire path of a random particle. If this leads to an overlap between the swapped path
and another path (i.e., two particles at the same time slice with a separation less than $a$),
then the other path is also swapped. This ``cascade'' continues until no overlaps remain.

Unfortunately, swap moves do not work as well as intended. As the simulation progresses,
Brownian bridge moves tend to nudge the particle paths into tight clusters near the well minima,
as noted above. Once the system is in that sort of configuration, a swap move has a high
probability of leading to a cascade that swaps every particle, which is equivalent to not swapping
any particle. This effect is worse for longer paths and larger $N$, and in practice,
the swap move was found to be mostly ineffective for the double well simulations described here.

\subsubsection{Potential moves}
Our ``potential moves'' were inspired by the parallel tempering technique~\cite{Swendsen1986}.
In parallel tempering, one runs multiple copies of a simulation at different temperatures
simultaneously, and exchanges configurations between two different simulations based
on the Metropolis criterion. This allows a simulation at a given temperature to sample a wider
variety of configurations, potentially avoiding an ergodicity problem.

In our potential moves, we run only one simulation, but we implement a move that changes the shape
of the external potential, specifically by changing $L$ from among a set of pre-defined values.
Given the current value of $L$, the potential move
attempts to change $L$ to the next highest or lowest value in the pre-defined set
and uses the Metropolis criterion to accept or reject the move.
The motivation here is to allow for a way to more easily change $(n_{l}-n_{r})^{2}$
for a high-barrier potential than would be possible with only Brownian bridge moves: lower
the potential barrier and then raise it again.

One challenge with this method is that certain potentials are more probable
than others (i.e., they have higher
average values of $\pi(X)$, where the average is taken over all configurations),
so a simulation with potential moves as described above would eventually end up only sampling the
most probable potential.  To avoid this problem, we introduce a set of weights, one per potential,
that we multiply by $\pi(x)$ before applying the Metropolis algorithm. We choose these weights
so that the average probability of transitioning from one potential to another is the same
as the probability of the reverse, which ensures that all of the potentials will be visited with
equal probability in the long run.  One can choose these weights using a version of the
Wang and Landau algorithm~\cite{Wang2001}.

In practice, these moves often work quite well once the correct weights are chosen.
However, there is still a problem: while weights can be chosen
to equalize the back-and-forth transition probabilities between two potentials, the actual
\emph{value} of that probability cannot be tuned at will and can be quite small.
If that is the case, then even though in principle all potentials will be visited
with equal frequency, that will only happen in practice in the limit of a very long simulation.
This situation arises for high-barrier potentials, and worsens for larger $N$ and longer paths;
see Table~\ref{tab:weightprobs} for an example.
\begin{table}
\begin{tabular}{rccc}
\toprule
& \multicolumn{3}{c}{Number of slices} \\
\cline{2-4}
$N$ & $100$ & $200$ & $800$\\
\colrule
8 & $4.8\times 10^{-1}$ & $3.9\times 10^{-1}$ & $6.1\times 10^{-2}$\\
16 & $3.0\times 10^{-1}$ & $1.5\times 10^{-1}$ & $1.8\times 10^{-2}$\\
32 & $1.2\times 10^{-1}$ & $4.6\times 10^{-2}$ & $1.0\times 10^{-5}$\\
64 & $4.8\times 10^{-2}$ & $1.1\times 10^{-2}$ & $3.1\times 10^{-12}$\\
\botrule
\end{tabular}
\caption{\label{tab:weightprobs}Probability of making a
``potential move'' that transitions between
two potentials of different shape, characterized by $L=2.875\,a_{ho}$ and $3\,a_{ho}$, for various
numbers of particles $N$ and path lengths (number of slices).  In all cases, $a=0.1\,a_{ho}$.}
\end{table}

\subsection{Off-diagonal observables}
\label{sec:offdiagonal}

The presentation of PIGS above describes the calculation of observables diagonal in the position
basis.  To compute an off-diagonal observable, such as the OBDM, we insert an extra set
of position eigenstates into Eq.~\eqref{eq:diagonalobs} at $R_{M+1}$:
\begin{equation}
\begin{split}
\braket{\hat{A}} = {} & \int \! dR_{0} \cdots dR_{2M+1} A(R_{M},R_{M+1}) \psi^{\ast}_{T}(R_{0}) \psi_{T}(R_{2M}) \\
& \times \prod_{i=0}^{M-1} G(R_{i}, R_{i+1}, \beta) \prod_{i=M+1}^{2M} G(R_{i}, R_{i+1}, \beta).
\end{split}
\end{equation}
There is no propagator connecting the configurations $M$ and $M+1$;
the path is said to be ``broken.''  The paths are sampled in
the same way as for diagonal observables, and the value of
the off-diagonal operator $\braket{\hat{A}}$
is estimated by averaging over $A(R_{M},R_{M+1})$ for the accepted paths,
just as before.

To compute the OBDM~\cite{Moroni2004}, the path of only one of the $N$ particles is broken
(i.e., $\textbf{r}_{M}$ is allowed to differ from $\textbf{r}_{M+1}$ for the broken path)
while $\textbf{r}_{M}$ is set equal to $\textbf{r}_{M+1}$ for the others.  One then samples paths
as usual and uses the set of accepted configurations to make a histogram of the occurrences
of particular pairs of $z_{M}$ and $z_{M+1}$ for the \emph{broken} path; this histogram
is $\rho(z,z^{\prime})$.
In order to normalize the OBDM, we multiply it by a factor such
that the sum of its eigenvalues (i.e., the total occupation of the natural orbitals) is 1.
Note that because of the finite bin sizes, this method can artificially generate non-physical
negative eigenvalues for small sampling, but these vanish given long enough simulations.

\section{Results \& Discussion}
\label{sec:results}

We now present the numerically exact PIGS results for squeezing and fragmentation of a BEC
in a three-dimensional double well potential, with a critical comparison to the corresponding
results from the two- and eight-mode approximations. We show results for the three double well
potentials with parameters $\alpha = 4/81\,a_{ho}^{-2}$ and $L=a_{ho}$, $2\,a_{ho}$,
and $3\,a_{ho}$.  As discussed above, these particular potentials are chosen to allow study
of a range of barrier heights while staying in a regime where it is sensible to apply both
the two-mode and eight-mode models.

\subsection{Two-mode model results}
\label{sec:2moderesults}
We first present the nearly-degenerate and exact two-mode model results for squeezing
and fragmentation, using the non-interacting one-body basis to allow analysis
of systematic trends over all three parameters $L$, $a$, and $N$.
\subsubsection{Squeezing}
\label{subsec:2moderesultssqueezing}
\begin{figure*}
\includegraphics{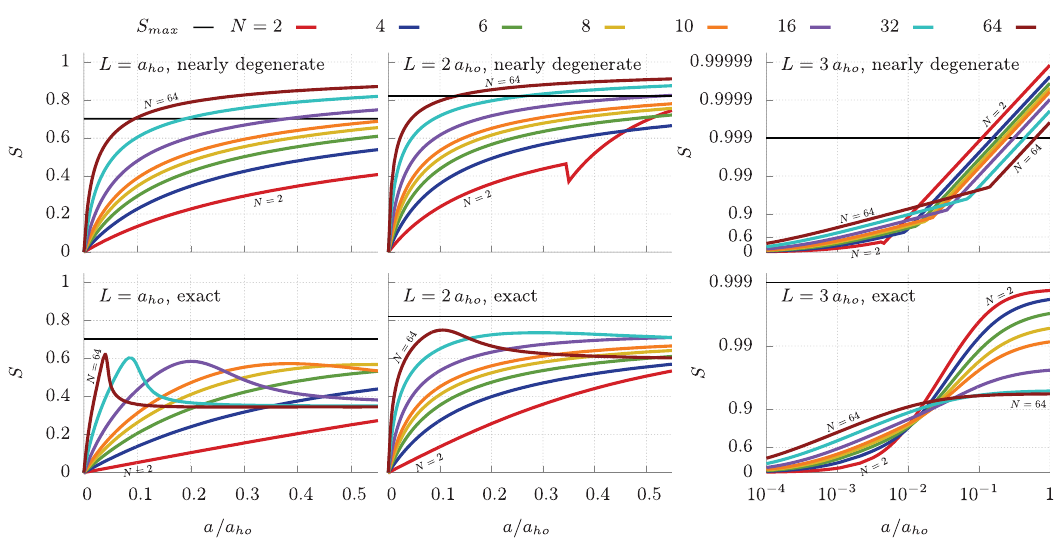}
\caption{\label{fig:2moderesults}(Color online) Squeezing $S$ vs. scattering length $a$ for
values of the particle number $N$ between 2 and 64, for three different  potentials
($L=a_{ho}$, $2\,a_{ho}$, and $3\,a_{ho}$ from left to right). The upper and lower plots
are of $S$ as computed in the nearly degenerate two-mode model (Eq.~\eqref{eq:JandIresult})
and the exact two-mode model (Eq.~\eqref{eq:squeezing2}), respectively.
All plots also indicate the value of $S_{max}$ from Eq.~\eqref{eq:Smax} by a horizontal line. For viewers of the figure in gray scale, labels in each panel indicate the plots for $N=2$ and 64; the other values of $N$ lie in between these in numerical order.}
\end{figure*}

Fig.~\ref{fig:2moderesults} shows the behavior of $S$ as a function of $a$ for a variety
of particle numbers and in three different double well geometries, for both the nearly degenerate
and the exact two-mode models. It is evident that the models agree well only for small values
of $a$, with deviations between them growing as $N$ increases.  There are several notable features
of the results in Fig.~\ref{fig:2moderesults} not seen in previous studies.
These are: (i) the lack of monotonicity (especially for low barriers) for the exact two-mode values
of $S$~vs.~$a$, (ii) the tendency of $S$ to increase as the extent of mode degeneracy increases
(i.e., for higher barriers), and (iii) the observed decrease with $N$ of the saturating $S$ value
at large $a$.

\begin{figure*}
\includegraphics{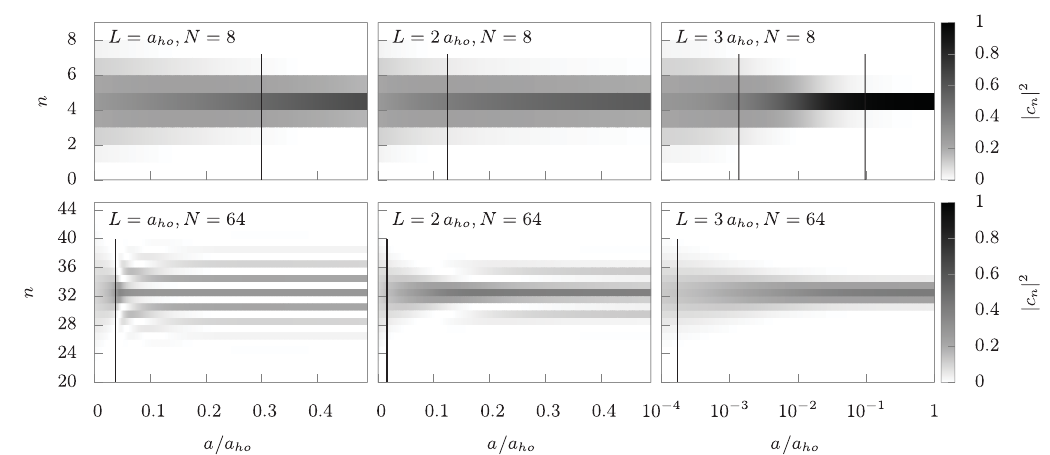}
\caption{\label{fig:groundstateVSa}
The components of the ground state $|c_{n}|^2 = |\braket{n|\psi_{ground}}|^2$ for the double well
potential as a function of scattering length $a$, as computed in the exact two-mode model
(Eqs.~\eqref{eq:2modeham_chi}--\eqref{eq:chi2}) for $N=8$ (top row) and 64 (bottom row)
for $L=a_{ho}$, $2\,a_{ho}$, and $3\,a_{ho}$ (left to right). The vertical lines indicate
the boundaries between the Rabi-like, Josephson-like, and Fock-like regions, from left to right in each plot
(although only the $L=3\,a_{ho}$, $N=8$ plot has a Fock-like regime visible).
In addition to the indicated regimes, the $L=a_{ho}$, $N=8$ plot has a Fock-like regime between
$a=9.03$ and 121 (and a Josephson-like regime thereafter) and the $L=a_{ho}$, $N=64$ plot
has a Fock-like regime between $a=1.85$ and 1.98 (and a Josephson-like regime thereafter).}
\end{figure*}

The degree of squeezing $S$ is of course a reflection of the composition of the ground state.
In Fig.~\ref{fig:groundstateVSa}, this composition is represented by plotting
$|c_{n}|^2 = |\braket{n|\psi_{ground}}|^2$ as a function of $a$.
The qualitative squeezing analysis in Sec.~\ref{sec:numdist} would suggest that each
of these plots should show
a smooth transition from a binomial distribution centered at $n=N/2$ to sole occupancy of the
$n=N/2$ state as $a$ increases from 0. This is clearly not what happens for $N=64$ particles.
For $L=a_{ho}$, the ground state settles into a very wide ``striped'' pattern, with occupancy
of every other Fock state, for $L=2\,a_{ho}$ it settles into a different striped pattern,
and for $L=3\,a_{ho}$ it settles into a narrow but wider-than-one-state distribution.
Even for 8 particles, the distribution narrows to the state $\ket{4}$ only for the highest barrier,
$L=3$, and the largest $a$ values, $a \geq 0.1$. Based on the definition of $S_{2}$,
Eq.~\eqref{eq:squeezing2}, the width or ``spread'' of these patterns gives a qualitative sense
of the degree of squeezing: narrower means more squeezing and vice versa. We can thereby see that
the progression of these patterns is consistent with the trends seen in the squeezing plots
displayed in Fig.~\ref{fig:2moderesults}.

Ideally, one could simply use the expression for $S$ in Eq.~\eqref{eq:squeezing2}
to explain the observed trends.  For example, the form of $S_{max}$ implies
that, in general, one should expect more squeezing for potentials with a higher degree
of degeneracy between the two modes; this is supported by the data. However, there is no explicit
analytical expression for the coefficients $c_{n}$ for the exact two-mode Hamiltonian that allows
the dependence of $S_2$ on $a$ and $L$ to be extracted.

We therefore study the ground state in different parameter regimes. In previous work,
the nearly-degenerate two-mode Hamiltonian, Eq.~\eqref{eq:JandIHam}, has been described in terms
of three regimes: Rabi, Josephson, and Fock (see, e.g.,~\cite{Leggett2001}).
The Rabi regime is the one in which the interwell interactions are negligible compared with
the effects of one-body tunneling, whereas in the other two regimes the interwell
interactions dominate.  The Josephson and Fock regimes are further distinguished in that the
interactions are so strong in the Fock regime that number fluctuations are suppressed
(i.e., the ground state is $\ket{N/2}$) while in the Josephson regime there are still
some fluctuations due to one-body tunneling.  These regimes are defined
by the value of the dimensionless parameter $\chi = 4a\kappa_{0}/\delta$,
which is the ratio of the interwell interaction and the one-body tunneling parameters
in Eq.~\eqref{eq:JandIHam}.  When this Hamiltonian is scaled to have a one-body tunneling
coefficient of unity, $\chi$ is the coefficient of the interwell (two-body) interaction term:
\begin{equation}
\hat{H} = -(\hat{a}_{l}^{\dag}\hat{a}_{r}+\hat{a}_{r}^{\dag}\hat{a}_{l}) - \chi \hat{n}_{l}\hat{n}_{r}.
\label{eq:JandIHam_scaled}
\end{equation}
Note that the one-body tunneling term scales like $N$ while the interwell interaction term
scales like $N^2$. This leads to an intuitive understanding of the definition of the Rabi regime
as $\chi \ll N^{-1}$, because for $\chi = N^{-1}$ the two terms are similar in size.
The Josephson regime is then given by $N^{-1} \ll \chi \ll N$ and the Fock regime by $N \ll \chi$;
at $\chi = N$ the second term in Eq.~\eqref{eq:JandIHam_scaled} is greater than the first
by a factor of roughly $N^2$.

To make contact between these three different regimes for the nearly-degenerate two-mode model
and the behavior of the exact two-mode system, we rewrite the full two-mode Hamiltonian
in the form
\begin{equation}
\hat{H} = - (\hat{a}_{l}^{\dag}\hat{a}_{r}+\hat{a}_{r}^{\dag}\hat{a}_{l}) + \chi_{1}(a,N)\hat{H}^{\prime},
\label{eq:2modeham_chi}
\end{equation}
where
\begin{align}
\hat{H}^{\prime}&= -\hat{n}_{l}\hat{n}_{r}+ \chi_{2} (\hat{a}_{l}^{\dag}\hat{a}_{l}^{\dag}\hat{a}_{r}\hat{a}_{r}+\hat{a}_{r}^{\dag}\hat{a}_{r}^{\dag}\hat{a}_{l}\hat{a}_{l}),
\label{eq:H'}\\
\chi_{1}(a,N) &= \frac{a\chi_{1}^{\ast}}{a^{\ast} - a(N-1)}, \label{eq:chi1}\\
\chi_{2} &= \frac{\kappa_{2}}{2(\kappa_{0}-2\kappa_{2})}, \label{eq:chi2}
\end{align}
and we have defined $a^{\ast} = \delta/4\kappa_{1}$ and
$\chi_{1}^{\ast} = (\kappa_{0}-2\kappa_{2})/\kappa_{1}$, both of which are
functions solely of the geometry of the double well.
Note that as in Eq.~\eqref{eq:JandIHam_scaled}, we have scaled the full two-mode Hamiltonian
to have unit amplitude of one-body tunneling.

We analyze the full two-mode Hamiltonian, Eq.~\eqref{eq:2modeham_chi}, in two stages.
First, we study the effect on the ground state of variations in $\chi_1$; for small $\chi_1$,
the Hamiltonian is dominated by one-body tunneling whereas for large $\chi_1$ it is dominated
by $\hat{H}^{\prime}$. Then, we study the effects of variations in $\chi_2$ on the ground state
of $\hat{H}^{\prime}$ alone (Fig.~\ref{fig:chi2plot}).  Finally, we combine these together to
understand the variation of both $\chi_1$ and $\chi_2$ on the ground state of the full Hamiltonian
Eq.~\eqref{eq:2modeham_chi} (Fig.~\ref{fig:chi1plots}).  For the geometries
studied in this work, $\kappa_{0}-2\kappa_{2}>0$ (see Table~\ref{tab:kappas}), so the signs
of $a^{\ast}$ and $\chi_{1}^{\ast}$ are the same as the sign of $\kappa_{1}$.
See Table~\ref{tab:chis} for representative numerical values for these parameters.
\begin{table}
\begin{tabular}{cccc}
\toprule
$L/a_{ho}$ & $a^{\ast}/a_{ho}$ & $\chi_{1}^{\ast}$ & $\chi_{2}$\\
\colrule
1 & $\,\,117.6$ & $\,\,\,\,\,\,\,48.2$ & $\,\,7.74\times 10^{-2}$ \\
2 & $-37.9$ & $\,\,\,-38.8$ & $\,\,3.01\times 10^{-2}$ \\
3 & $\,\,\,-6.0$ & $-553.2$ & $\,\,9.73\times 10^{-6}$ \\
\botrule
\end{tabular}
\caption{\label{tab:chis} Values of $a^{\ast}$, $\chi_{1}^{\ast}$, and $\chi_{2}$
for $L=a_{ho}$, $2\,a_{ho}$, and $3\,a_{ho}$.}
\end{table}

In the nearly degenerate limit, both $\kappa_{1}$ and $\kappa_{2}$ equal $0$; therefore, $\chi_1$ reduces to $\chi$, $\chi_2$ to $0$, and Eq.~\eqref{eq:2modeham_chi} to Eq.~\eqref{eq:JandIHam_scaled} (as it should). This motivates us to generalize the definitions
of the three two-mode regimes that have been defined previously for the nearly-degenerate
two-mode model, to the full two-mode system.  As an example, recall that physically the Rabi regime
is the one in which interwell interactions are negligible.  Because $\chi_{1}$ can be both positive
or negative, this physical condition corresponds mathematically to both $0< \chi_{1} \ll N^{-1}$ and
$0 > \chi_{1} \gg -N^{-1}$ (or equivalently, to the single condition $|\chi_{1}| \ll N^{-1}$)
in Eq.~\eqref{eq:2modeham_chi}. Hence, we define a Rabi-like regime by $|\chi_{1}| \ll N^{-1}$.
Similarly, we define a Josephson-like regime by $N^{-1} \ll |\chi_{1}| \ll N$ and a Fock-like regime
by $N \ll |\chi_{1}|$.  In contrast, $\chi$ is always positive for repulsive interactions
(i.e., when $a>0$), so no absolute value signs are needed in the inequalities
after Eq.~\eqref{eq:JandIHam_scaled}.

\begin{figure}
\includegraphics{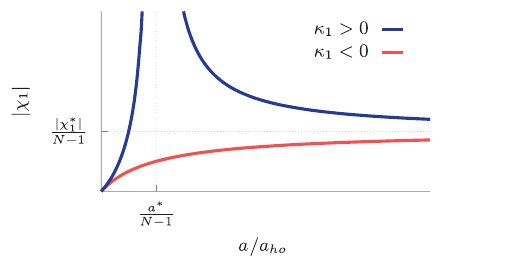}
\caption{(Color online) Schematic plot of $|\chi_{1}|$ as a function of $a$ for both
positive and negative $\kappa_{1}$.  For positive $\kappa_{1}$, $|\chi_{1}|$ diverges
at $a^{\ast}/(N-1)$ and then asymptotes to $|\chi_{1}^{\ast}|/(N-1)$ from above;
all three regimes (Rabi-like, Josephson-like, and Fock-like) are present for some range of $a$.
For negative $\kappa_{1}$, $|\chi_{1}|$ asymptotes to $|\chi_{1}^{\ast}|/(N-1)$ from below,
and there may or may not be Josephson-like or Fock-like regimes depending on the size
of $|\chi_{1}^{\ast}|/(N-1)$ compared with $N$ and $N^{-1}$.
}
\label{fig:chi1}
\end{figure}

\begin{figure*}
\includegraphics{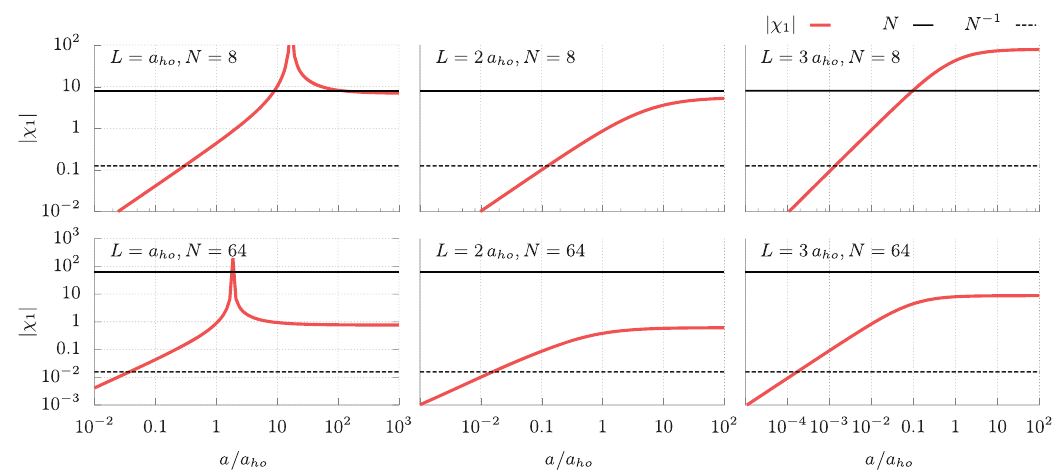}
\caption
{\label{fig:chi1cases} (Color online) The value of the parameter $|\chi_{1}|$ from
the full two-mode Hamiltonian~(Eq.~\eqref{eq:chi1}) as a function of $a$, for $L=a_{ho}$,
$2\,a_{ho}$, and $3\,a_{ho}$ and $N=8$ and $64$. Also indicated in each plot are the values of
$N$ (solid horizontal line) and $N^{-1}$ (dotted horizontal line).  The system is in
a Rabi-like regime when $|\chi_{1}|$ is below the dotted line, in a Josephson-like regime
when $|\chi_{1}|$ is between the two lines, and in a Fock-like regime when $|\chi_{1}|$ is above
the solid line. We see that while the Rabi-like regime must always be present
(since $\chi_{1} = 0$ for $a=0$), the other two need not be.}
\end{figure*}

\begin{figure}
\includegraphics{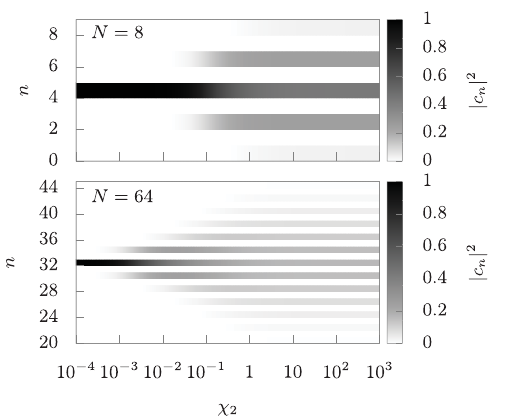}
\caption{\label{fig:chi2plot} The components of the ground state of the Hamiltonian $H^{\prime}$,
Eq.~\eqref{eq:H'}, as a function of $\chi_{2}$ for $N=8$ and 64.
True Rabi, Josephson, and Fock regimes are only possible when $\chi_{2}$ is small,
i.e., when the ground state is dominated by the Fock state $\ket{N/2}$.
}
\end{figure}

\begin{figure*}
\includegraphics{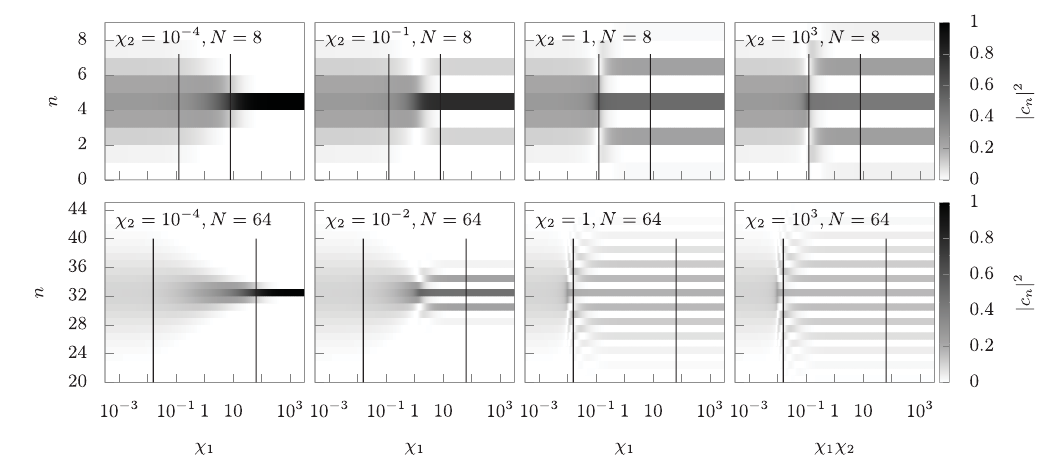}
\caption{\label{fig:chi1plots}The components of the ground state of the full two-mode Hamiltonian
(Eq.~\eqref{eq:2modeham_chi}) for $N=8$ and 64 and a range of $\chi_{2}$ values. The leftmost three columns are plotted as a function of $\chi_{1}$, while the rightmost column is plotted as a function of the product $\chi_1\chi_2$ because this is the column for which $\chi_2 >1$ (see p.~\pageref{text:chi1VSchi1chi2} for a discussion of this distinction). The vertical lines indicate
the boundaries between the Rabi-like, Josephson-like, and Fock-like regimes, from left to right in each plot.}
\end{figure*}

In Fig.~\ref{fig:chi1} we schematically plot $|\chi_{1}|$ as a function of $a$ for both
positive and negative $\kappa_{1}$.  For positive $\kappa_{1}$, we see that the system will have
Rabi-like, Josephson-like, and Fock-like regimes for some range of interaction strength $a$,
since $|\chi_{1}|$ diverges at $a = a^{\ast}/(N-1)$.  For negative $\kappa_{1}$,
the system will have a Rabi-like regime but may or may not have Josephson-like or Fock-like regimes,
depending on whether $|\chi_{1}^{\ast}| < N^{-1}$, $N^{-1} < |\chi_{1}^{\ast}| < N$,
or $N < |\chi_{1}^{\ast}|$.  In Fig.~\ref{fig:chi1cases}, we plot $|\chi_{1}|$ as a function
of $a$ for the same six sets of parameters that are depicted in Fig.~\ref{fig:groundstateVSa};
we also include the values of $N$ and $N^{-1}$ in the plots to make it clear where transitions
between the three regimes occur.  These transitions correspond to the vertical lines
in Fig.~\ref{fig:groundstateVSa} (see also the description in the caption
to Fig.~\ref{fig:groundstateVSa}).

In the case where $\chi_{1}$ is large and the Hamiltonian is dominated by $\hat{H}^{\prime}$,
the ground state also depends on the value of $\chi_2$. In Fig.~\ref{fig:chi2plot}, we plot
the ground state of $\hat{H}^{\prime}$ alone as a function of $\chi_{2}$ for $N=8$ and 64.
The ground state progresses from $\ket{N/2}$ to a striped pattern as $\chi_{2}$ increases.
This comes about because $\hat{H}^{\prime}$ can be rewritten in a block-diagonal form
with two tridiagonal blocks, where each block involves either the even-numbered or the odd-numbered
Fock states and therefore the ground state has contributions from only one of these.
Inspection of Eq.~\eqref{eq:H'} shows that the interwell interaction term
($-\hat{n}_{l}\hat{n}_{r}$) dominates when $\chi_{2} \ll 1$ while the two-body tunneling term
dominates when $\chi_{2} \gg 1$ (since both terms of Eq.~\eqref{eq:H'} scale like $N^2$, the definitions of these regimes must be independent of $N$; this is consistent with Fig.~\ref{fig:chi2plot}).
Thus, true Rabi, Josephson, and Fock regimes for the full Hamiltonian are only possible when $\chi_2$ is very small; when $\chi_2$ is large, only Rabi-like, Josephson-like, and Fock-like regimes are possible.
Also, when $\chi_{2}$ is large there is an additional subtlety to be taken into account.
The identification of $\chi_{1}$ as the parameter that distinguishes the three regimes depends
on the two terms in $\hat{H}^{\prime}$ having coefficients
less than or equal to 1; this does not hold for large $\chi_{2}$. In that case,
one should pull the factor of $\chi_{2}$ out of $\hat{H}^{\prime}$, and the transitions among
Rabi-like, Josephson-like, and Fock-like regimes are defined instead by the size of the product $\chi_{1}\chi_{2}$. \label{text:chi1VSchi1chi2}

Given this interpretation of $\chi_{1}$ and $\chi_{2}$, we can now understand the patterns in
Fig.~\ref{fig:chi1plots}, where the ground state of the full two-mode Hamiltonian
(Eq.~\eqref{eq:2modeham_chi}) is plotted as a function of $\chi_{1}$ for various values of $\chi_{2}$
for $N=8$ and 64. In the Rabi-like regime ($\chi_{1} << N^{-1}$), the ground state is close
to the binomial distribution of the one-body tunneling terms regardless of the size of $\chi_{2}$.
In the Fock-like regime, ($\chi_{1} >> N$), the ground state varies from $\ket{N/2}$
to a wide striped pattern as $\chi_{2}$ increases.  The Josephson-like regime interpolates between
the other two, with a narrow ``neck'' where the binomial and striped patters touch.
Recall that the squeezing parameter $S$ varies like the width of these distributions
(recall (Eq.~\eqref{eq:squeezing2}), so the neck corresponds to a peak in $S$.
The rightmost panel shows the case when $\chi_2 > 1$, where the three regimes are now determined by the
product $\chi_1 \chi_2$ rather than by $\chi_1$ alone.

With this understanding, we can now return to the trends in two-mode data presented
in Figs.~\ref{fig:2moderesults} and \ref{fig:groundstateVSa} and provide a detailed interpretation.
The closer the system is to the degenerate two-mode case (i.e., larger $L$, smaller $\kappa_{2}$,
and smaller $\chi_{2}$), the closer the ground state will be to $\ket{N/2}$ (as opposed to
a striped state) for large $a$.  This implies that there will generally be more squeezing
with increased degeneracy.  Likewise, the closer the system is to the degenerate limit,
the more likely that $S$ varies monotonically with $a$: because the large-$a$ state is narrower,
the neck in the Josephson-like regime (and hence the peak in $S$) will be less pronounced
or non-existent (upper half and rightmost plot of lower half of Figs.~\ref{fig:2moderesults}
and~\ref{fig:groundstateVSa}).  For small values of $a$, increasing $N$ tends to increase squeezing,
while this is not always true for large $a$.  This change in behavior can be understood by noting
that for very large $a$, $\chi_{1} = \chi_{1}^{\ast}/(N-1)$.  This quantity decrease
as $N$ increases, and therefore the Hamiltonian becomes increasingly dominated by the one-body
tunneling terms, which have a wide distribution of Fock states in the ground state.
Hence, the squeezing $S$ is expected to \emph{decrease} with increasing $N$ for very large $a$:
this is confirmed by the plots in the righthand panels of Figs.~\ref{fig:2moderesults}
and \ref{fig:groundstateVSa}.

We can compare a small subset of these squeezing results with the the results of the exact two-mode
calculations in~\cite{Spekkens1999}, obtained for a double well potential with a form that
is comparable (but not identical) to ours.  For a given barrier height centered at $z=0$,
the potential minima in Ref.~\cite{Spekkens1999} are closer than ours by approximately a factor of 2
(e.g., our potential with $L=2\, a_{ho}$ has barrier height $0.395\, E_{ho}$ and minima
at $2\, a_{ho}$, while the potential characterized by $\alpha = 15\,a_{ho}E_{ho}$
in~\cite{Spekkens1999} has barrier height $0.339\, E_{ho}$ and minima at $1.16\, a_{ho}$).
Ref.~\cite{Spekkens1999} reports calculations of the quantity $\Delta N_{1} = \sqrt{N(1-S)}/2$
computed for the exact two-mode Hamiltonian with $N=100$ particles and scattering length
$a = 6.24 \times 10^{-4} \, a_{ho}$, and found that $\Delta N_{1}$ decreased as
the barrier height increased.  In Table~\ref{tab:SandS}, we give the corresponding values
of $\Delta N_{1}$ for our exact two-mode calculations, with $N=64$ particles and scattering
length $a = 6.31 \times 10^{-4} \, a_{ho}$.  It is apparent that $\Delta N_{1}$ decreases
with increasing $L$ (increasing barrier height), consistent with Figure 2 in~\cite{Spekkens1999}.
\begin{table}
\begin{tabular}{ccc}
\toprule
$L/a_{ho}$ & $\mathcal{C}^{(1)}$ & $\Delta N_{1}$\\
\colrule
1 & 0.999993 & 3.959 \\
2 & 0.999967 & 3.897 \\
3 & 0.986015 & 2.138 \\
\botrule
\end{tabular}
\caption{\label{tab:SandS} Values of squeezing measure $\Delta N_{1}$ and fragmentation measure $\mathcal{C}^{(1)}$ defined in~\cite{Spekkens1999}, calculated for the exact two-mode model
with $N=64$ particles, scattering length $a = 6.31 \times 10^{-4} \, a_{ho}$, and the potential barrier parameter $L$ taking on values $a_{ho}$, $2\,a_{ho}$, and $3\,a_{ho}$.}
\end{table}

\subsubsection{Fragmentation and Depletion}
\label{subsub:2FD}

\begin{figure*}
\includegraphics{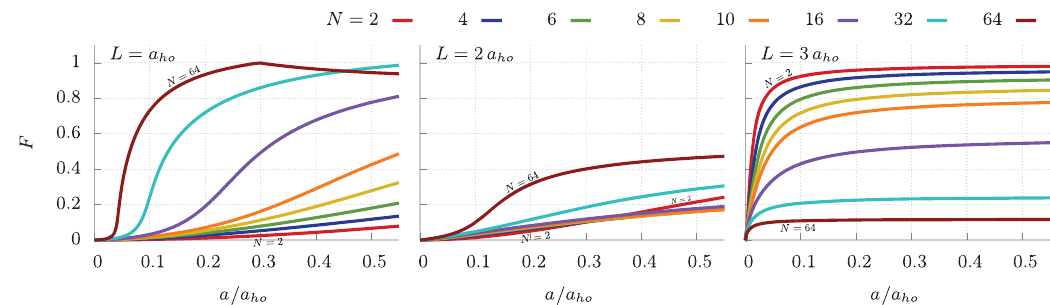}
\caption{\label{fig:2modefrag}(Color online) Fragmentation $F$ as a function of scattering length
$a$, for values of particle number $N$ between 2 and 64 for three different potentials ($L=a_{ho}$,
$2\,a_{ho}$, and $3\,a_{ho}$, from left to right) in the exact two-mode model. Because there are
only two modes in this model, the depletion parameter $D$ is necessarily zero. For viewers of the figure in gray scale, labels in each panel indicate the plots for $N=2$ and 64; the other values of $N$ lie in between these in numerical order.}
\end{figure*}

Fig.~\ref{fig:2modefrag} shows the fragmentation parameter $F$ as a function of $a$
for the three double well geometries in the full two-mode description.  Recall that the larger
the value of $F$, the more fragmentation in the system. Fig.~\ref{fig:2modefrag} shows that
in general, fragmentation increases with $a$.  An exception occurs for large $a$ when $N=64$
and $L=a_{ho}$, for which $\mathcal{N}_{1}$ becomes greater than $\mathcal{N}_{0}$, and therefore
the amount of fragmentation drops
after reaching a maximum of 1.  However, the most notable feature of these plots
is that for low barriers ($L=a_{ho}$), systems with larger $N$ experience much more fragmentation
than systems with smaller $N$, whereas the opposite is true for systems with high barriers
($L=3\,a_{ho}$). Equivalently, for small $N$, increasing the barrier height increases the amount
of fragmentation, while for large $N$, increasing the barrier height decreases the amount
of fragmentation.

We can understand these trends by examining the structure of the ground state
revealed in Fig.~\ref{fig:groundstateVSa}.  First, consider the high barrier limit
($L=3\,a_{ho}$). Based on the arguments above, we expect the ground state to be narrower
(closer to $\ket{N/2}$) for smaller $N$.  From the analytic form of $F$ given in Eq.~\eqref{eq:F2},
we see that the terms in the sum depend on $c_{n}c_{n+1}$, i.e. the product of two adjacent
$c_{n}$'s; the smaller the products, the more fragmentation there is.  In general, then, we expect
a narrower ground state to have more fragmentation, because it will have smaller $c_{n}c_{n+1}$
products than a wider ground state (indeed, if the ground state is $\ket{N/2}$, all of the
$c_{n}c_{n+1}$ products are zero).
Hence for $L=3\,a_{ho}$, fragmentation decreases with $N$ at large $a$.

For low barriers ($L=a_{ho}$), the situation is different.  As $N$ increases, the ground state
widens (as with $L=3\,a_{ho}$), but it does so by developing a striped pattern. Despite the fact
that the pattern is wide for large $N$, the striping will cause the $c_{n}c_{n+1}$ products
in Eq.~\eqref{eq:F2} to be small, because for each pair of adjacent $c_{n}$'s, one of them
will be close to zero. Hence for $L=a_{ho}$
at large $a$ values, there will be more fragmentation for large $N$ than for
small $N$. The fragmentation pattern in the intermediate barrier regime
($L=2\,a_{ho}$) is a crossover between the low and high barrier situations.

To connect these ideas back to the structure of the full two-mode Hamiltonian given
in Eq.~\eqref{eq:2modeham}, recall that the degree of striping observed in the ground state
is determined by $\chi_{2}$, which controls the strength of the two-body tunneling terms
$\hat{a}_{l}^{\dag}\hat{a}_{l}^{\dag}\hat{a}_{r}\hat{a}_{r}+
\hat{a}_{r}^{\dag}\hat{a}_{r}^{\dag}\hat{a}_{l}\hat{a}_{l}$ (see Eq.~\eqref{eq:H'}).
For high barriers $\chi_{2}$ is small and these terms are negligible, and vice versa for
low barriers. The authors of~\cite{Bader2009} similarly found that two-body tunneling terms
are critical to the onset of fragmentation of bosons in a single well potential,
with a study of $N=100$ particles showing that fragmentation increases as the strength
of the two-body tunneling terms increases.  This is consistent with Fig.~\ref{fig:2modefrag},
where the fragmentation $F$ for the largest $N$ value ($N=64$) is seen to increase as $L$ decreases;
Table~\ref{tab:chis} shows that decreasing $L$ is a proxy for increasing $\chi_{2}$, which controls
the strength of the two-body tunneling terms in Eq.~\eqref{eq:chi2}.  This is an example
of the general phenomenon of interaction-induced fragmentation
due to pair exchanges~\cite{Fischer2010,Bader2013}.

Finally, as with the squeezing results, we may compare a subset of our fragmentation results
with the corresponding results in~\cite{Spekkens1999}.  For fragmentation, the relevant parameter
to compare is $\mathcal{C}^{(1)} = 1-F$.
Table~\ref{tab:SandS} shows that this parameter decreases
as the height of the double well barrier increases, also consistent with the findings
in~\cite{Spekkens1999}.

\subsubsection{Summary}

Previous two-mode studies, often conducted with a restricted two-mode model that is relevant
only when the barrier is strong and therefore and the modes nearly degenerate, have predicted
that squeezing should monotonically increase with $a$ and that fragmentation should monotonically
increase with barrier strength. Instead, by including all possible contributions to the two-mode
Hamiltonian, we find a much richer behavior, with the following characteristics:
\begin{itemize}
\item Squeezing is not necessarily monotonic with $a$, especially for weak barriers.
\item For a given $N$, squeezing tends to increase with barrier strength.
\item For a given barrier strength, squeezing tends to decrease with $N$ for large $a$.
\item For fixed $a$, fragmentation tends to increase with $N$ for weak barriers,
whereas fragmentation tends to decrease with $N$ for strong barriers.
\end{itemize}

These trends are explained above by understanding how $a$, $N$, and the double well geometry
parameters influence the relative importance of the terms in the two-mode Hamiltonian,
and therefore change the nature of the ground state.  The terms in the Hamiltonian come in three
types:
\begin{enumerate}
\item Terms that involve a single Fock state. The ground state of these terms considered alone
is $\ket{N/2}$, which exhibits high squeezing and high fragmentation.
\item Terms that involve transitions between Fock states that involve a single particle.
The ground state of these terms considered alone is a mix of states binomially distributed
around $\ket{N/2}$, which exhibits low squeezing and low fragmentation.
\item Terms that involve transitions between Fock states that involve two particles.
The ground state of these terms considered alone is a mix of \emph{alternating} states
distributed around $\ket{N/2}$ (i.e., it includes $\ket{N/2}$, $\ket{N/2 \pm 2}$,
$\ket{N/2 \pm 4}$, etc.), which exhibits high squeezing and low fragmentation.
\end{enumerate}
When interactions are weak ($|\chi_{1}| \ll N^{-1}$), type 2 terms dominate regardless
of the strength of the barrier (Rabi-like regime). When interactions are strong
($|\chi_{1}| \gg N$), the strength of the barrier matters (Fock-like regime): for low barriers
(large $\chi_{2}$), type 3 terms dominate, while for high barriers (small $\chi_{2}$),
type 1 terms dominate.  The Josephson-like regime interpolates between these two regimes
and is characterized by $N^{-1} \ll |\chi_{1}| \ll N$.

\subsection{Eight-mode model results}
\label{sec:8moderesults}
We now present the eight-mode model results for squeezing and fragmentation, calculated with the non-interacting one-body basis.
\subsubsection{Squeezing}
\begin{figure*}
\includegraphics{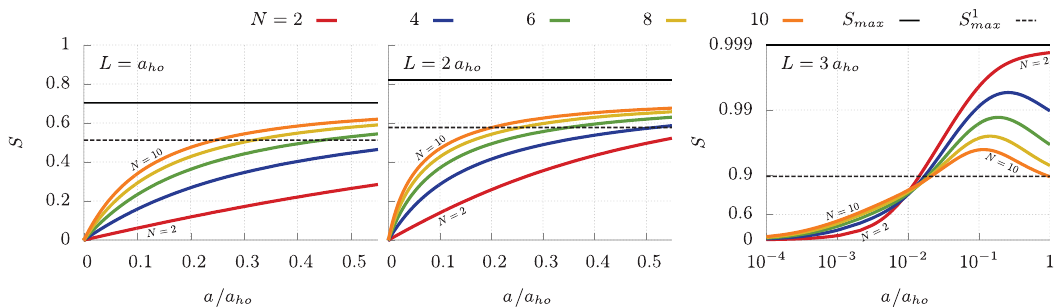}
\caption{\label{fig:8moderesults}(Color online) Squeezing $S$ as a function of scattering length $a$
for values of particle number $N$ between 2 and 10, for three different potentials ($L=a_{ho}$,
$2\,a_{ho}$, and $3\,a_{ho}$, from left to right) in the eight-mode model. The horizontal lines
indicate the values of $S_{max}$ and $S^{1}_{max}$. For viewers of the figure in gray scale, labels in each panel indicate the plots for $N=2$ and 10; the other values of $N$ lie in between these in numerical order.}
\end{figure*}

Fig.~\ref{fig:8moderesults} shows the behavior of $S$ for the eight-mode model as a function of $a$,
for $N \leq 10$.  The squeezing behavior for the low barriers ($L=a_{ho}$ and $2\,a_{ho}$)
looks qualitatively similar to the corresponding small $N$ results ($N \leq 10$) for the nearly
degenerate and exact two-mode models in Fig.~\ref{fig:2moderesults}, showing a monotonic increase
with interaction strength $a$. However, for $L=3\,a_{ho}$ we see qualitatively different behavior:
here the squeezing shows distinctly non-monotonic behavior, with a clear maximum that moves
to smaller values of $a$ for larger $N$ values.  We would like to account for this behavior,
despite the fact that the eight-mode Hamiltonian cannot be analytically analyzed as easily
as the two-mode Hamiltonian because of its complexity.

\begin{figure*}
\includegraphics{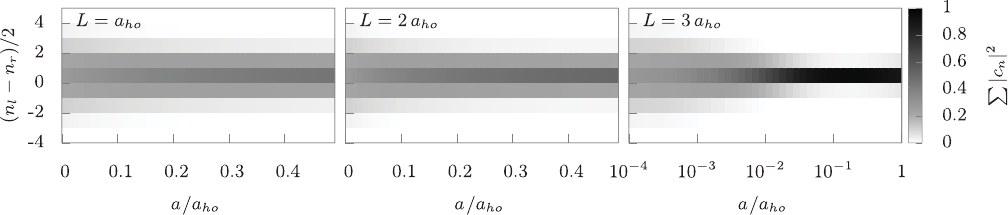}
\caption{\label{fig:8modegroundstateVSa}
Representation of the ground state for the double well potential as a function of scattering length
$a$, as computed in the eight-mode model for $N=8$ and various $L$ ($a_{ho}$, $2\,a_{ho}$,
and $3\,a_{ho}$, from left to right). The quantity plotted in greyscale is the sum of $|c_{n}|^{2}$
for components of the ground state with a given value for the difference between the number
of particles in left and right modes, $n_l - n_r$ ($y$-axis).}
\end{figure*}

Since the Hilbert space for the eight-mode model is so large, it is not useful to plot
the contribution to the ground state of each individual Fock state as in
Fig.~\ref{fig:groundstateVSa}. Instead, we sum the contributions of all Fock states for which
the difference in the number of particles occupying left and right modes is the same,
regardless  of which exact modes are occupied; this representation is shown in
Fig.~\ref{fig:8modegroundstateVSa}. By comparing Fig.~\ref{fig:8moderesults} and
Fig.~\ref{fig:8modegroundstateVSa}, we see that, just as in the two-mode model,
a narrower distribution corresponds to more squeezing and that the maximum in the $L=3\,a_{ho}$
squeezing data corresponds to a narrow neck in the state distribution.

It is also useful to consider the implications of the eight-mode analog to the quantity
$S_{max}$, Eq.~\eqref{eq:Smax}, which puts a constraint on the maximum amount of squeezing
possible for the two-mode model in a given double well potential.  Recall that $S_{max}$ measures
the fraction of a left or right localized state on the ``wrong'' side of $z=0$;
the more ``spillover,'' the smaller $S_{max}$.   In general, the modes that involve $\phi_{L/R}(z)$
(with $nlm = 210$) will have more spillover than the other six states, which all involve
$\phi_{l/r}(z)$ (see Eq.~\eqref{eq:eightmode_modes} and compare Figs.~\ref{fig:psi_2}
and \ref{fig:psi_8}). This motivates us to define $S^{1}_{max}$,
\begin{equation}
S^{1}_{max} = \left(1- 2\int_{-\infty}^{0} |\phi_{L}(z)|^{2} \,dz \right)^{2},
\label{eq:S1max}
\end{equation}
as a measure of the spillover of the $nlm=210$ modes; this quantity is analogous to $S_{max}$
for the $nlm=100$ and 21$\pm$1 modes. For $L=a_{ho}$, $2\,a_{ho}$, and $3\,a_{ho}$,
$S_{max} = 0.704$, 0.820, and 0.999 and $S^{1}_{max} = 0.512$,
0.577, and 0.895, respectively (these values are included in Fig.~\ref{fig:8moderesults}).
Both $S_{max}$ and $S^{1}_{max}$ increase with $L$, so there is more potential for squeezing
for higher barriers.  However, note that $S^{1}_{max} < S_{max}$ for each $L$:
this implies that the 210 modes have less potential for squeezing than the other modes.
Hence, by analogy with the exact two-mode analysis (see discussion after Eq.~\eqref{eq:Smax}),
we expect that a ground state will have less squeezing, other things being equal,
if it is dominated by the 210 modes than if it is dominated by the others. In other words,
Fock states with a given difference between the number of particles occupying left and right modes
will contribute less to squeezing if they are dominated by the 210 modes because these modes
have more spillover across $z=0$.

\begin{figure*}
\includegraphics{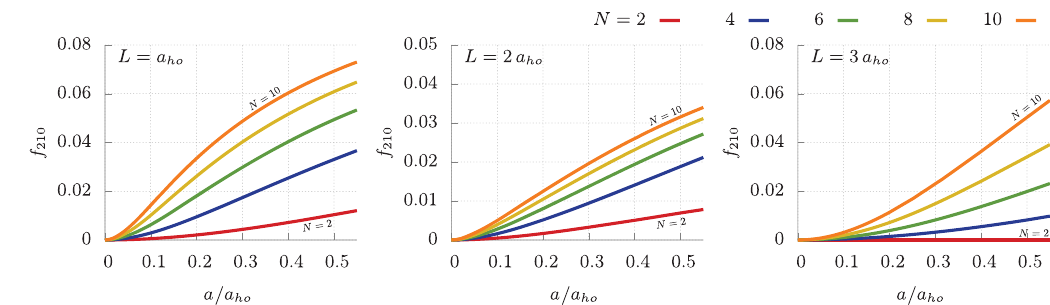}
\caption{\label{fig:8modeoccupation}(Color online) Fraction $f_{210}$ of particles in the 210 modes
of the eight-mode model as a function of the scattering length $a$, for various particle number
$N$ between 2 and 10 and for three different potentials ($L=a_{ho}$, $2\,a_{ho}$, and $3\,a_{ho}$,
from left to right).  The total fraction of particles in the 100 and 21$\pm$1 modes (not shown)
is given by $1-f$. For viewers of the figure in gray scale, labels in each panel indicate the plots for $N=2$ and 10; the other values of $N$ lie in between these in numerical order.}
\end{figure*}

Fig.~\ref{fig:8modeoccupation} shows the fraction of particles in the eight-mode model ground state
that are in the $210$ modes. For a given value of $a$, that fraction increases with $N$,
reaching as high as 8 percent for the largest $N$ values.  This is reasonable: in general,
the repulsive interaction between the particles drives them apart, and in a three-dimensional
eight-mode model, one way that the particles can avoid each other is by occupying modes with different
values of $m$. Hence, we expect that, for a given $a$, increasing the number of particles $N$
will result in a larger fraction occupying the $210$ modes, which are more strongly delocalized
than the other modes.  Thus, the occupation of these modes can then reduce the amount of
squeezing via the spillover mechanism described above.

\subsubsection{Fragmentation and Depletion}

\begin{figure*}
\includegraphics{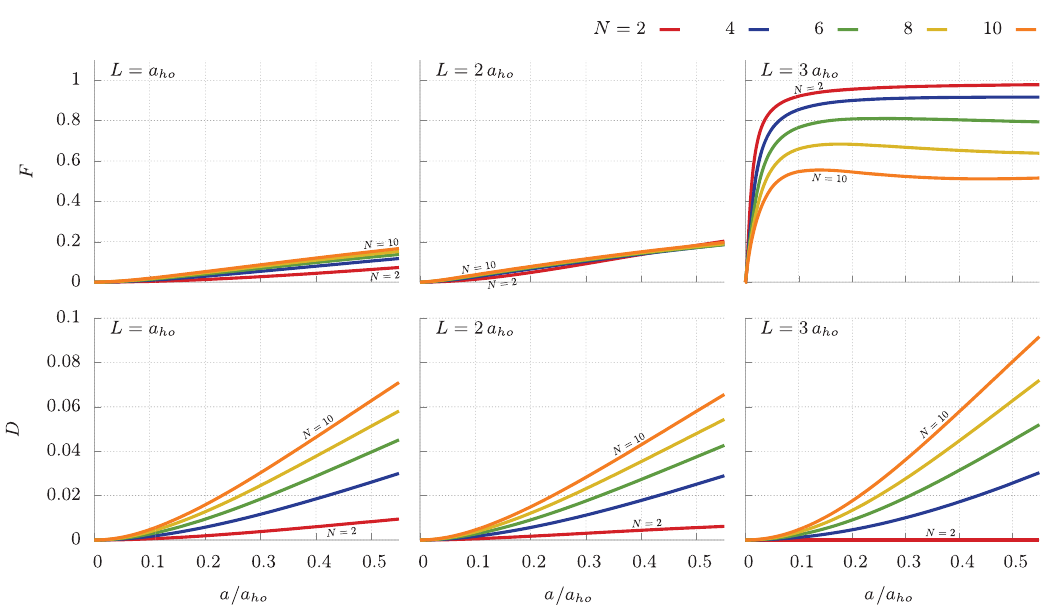}
\caption[Degree of fragmentation and depletion vs.~scattering length in the eight-mode model]
{\label{fig:8modefrag} (Color online)
Fragmentation $F$ (top) and depletion $D$ (bottom) as a function of~scattering length $a$
for values of particle number $N$ between 2 and 10, for three different potentials
($L=a_{ho}$, $2\,a_{ho}$, and $3\,a_{ho}$, from left to right) in the eight-mode model. For viewers of the figure in gray scale, labels in each panel indicate the plots for $N=2$ and 10; the other values of $N$ lie in between these in numerical order.}
\end{figure*}

Fig.~\ref{fig:8modefrag} shows the fragmentation and depletion parameters $F$ and $D$
(Eqs.~\eqref{eq:F} and \eqref{eq:D}), as a function of scattering length $a$ in the eight-mode model.
The main qualitative differences between these predictions and those of the two-mode model
in Fig.~\ref{fig:2modefrag} are that the eight-mode states exhibit significantly less fragmentation
and there is now nonzero depletion. In the case of the largest barrier height $L=3$,
we also now find non-monotonic dependence of $F$ on $a$ for the largest particle number,
$N =10$ (upper right hand panel).

In the eight-mode model, there are eight natural orbitals to occupy, rather than two.
The fact that there is less fragmentation and simultaneously now also depletion in the eight-mode
case implies that the occupation of the orbitals is spread out among more than just the first two,
but also that $\mathcal{N}_{0}$ is larger relative to $\mathcal{N}_{1}$ in the eight-mode case
than in the two-mode case. When there are only two modes, the only way to reduce $\mathcal{N}_{0}$
is to increase $\mathcal{N}_{1}$. However, because there are six other natural orbitals to occupy
in the eight-mode model, a reduction in $\mathcal{N}_{0}$ can be compensated by an increase
in any of $\mathcal{N}_{1}$ through $\mathcal{N}_{7}$.  Hence, conditions that would have led
to pure fragmentation in the two-mode case lead to less fragmentation with some depletion
in the eight-mode case.

\subsubsection{Summary}

Compared with the two-mode ground state, we find that the eight-mode ground state
\begin{itemize}
\item exhibits less squeezing, especially for $L=3\,a_{ho}$,
\item exhibits a maximum in $S$ vs. $a$ for $L=3\,a_{ho}$, and
\item exhibits less fragmentation and more depletion.
\end{itemize}
Physically, these effects can be understood as a consequence of the occupation of modes
in the $n=2$ energy level in addition to the two-mode model's $n=1$ modes. Since they are
more delocalized across the barrier, the $n=2$ modes lead to greater particle fluctuations
for the same value of $N_{l} - N_{r}$ than do the $n=1$ modes.  Hence, to the extent that they
are occupied in the ground state, those $n=2$ modes will tend to suppress both squeezing
and fragmentation in the eight-mode ground states, relative to two-mode ground states.
The presence of a maximum in the squeezing parameter comes about because of a competition between
this effect and the usual suppression of particle fluctuations that comes about for increased
scattering length $a$.  Finally, the presence of more than two natural orbitals in the eight-mode
model allows for non-zero depletion when the ``extra'' modes are occupied, while depletion is zero
by definition in the two-mode model.

\subsection{Quantum Monte Carlo results}
We now present the numerically exact PIGS results for squeezing and fragmentation, and compare
these to the results from the truncated basis calculations within the two- and eight-mode models.

\subsubsection{Squeezing}
\label{subsub:QMCsqueezingresults}
\begin{figure*}
\includegraphics{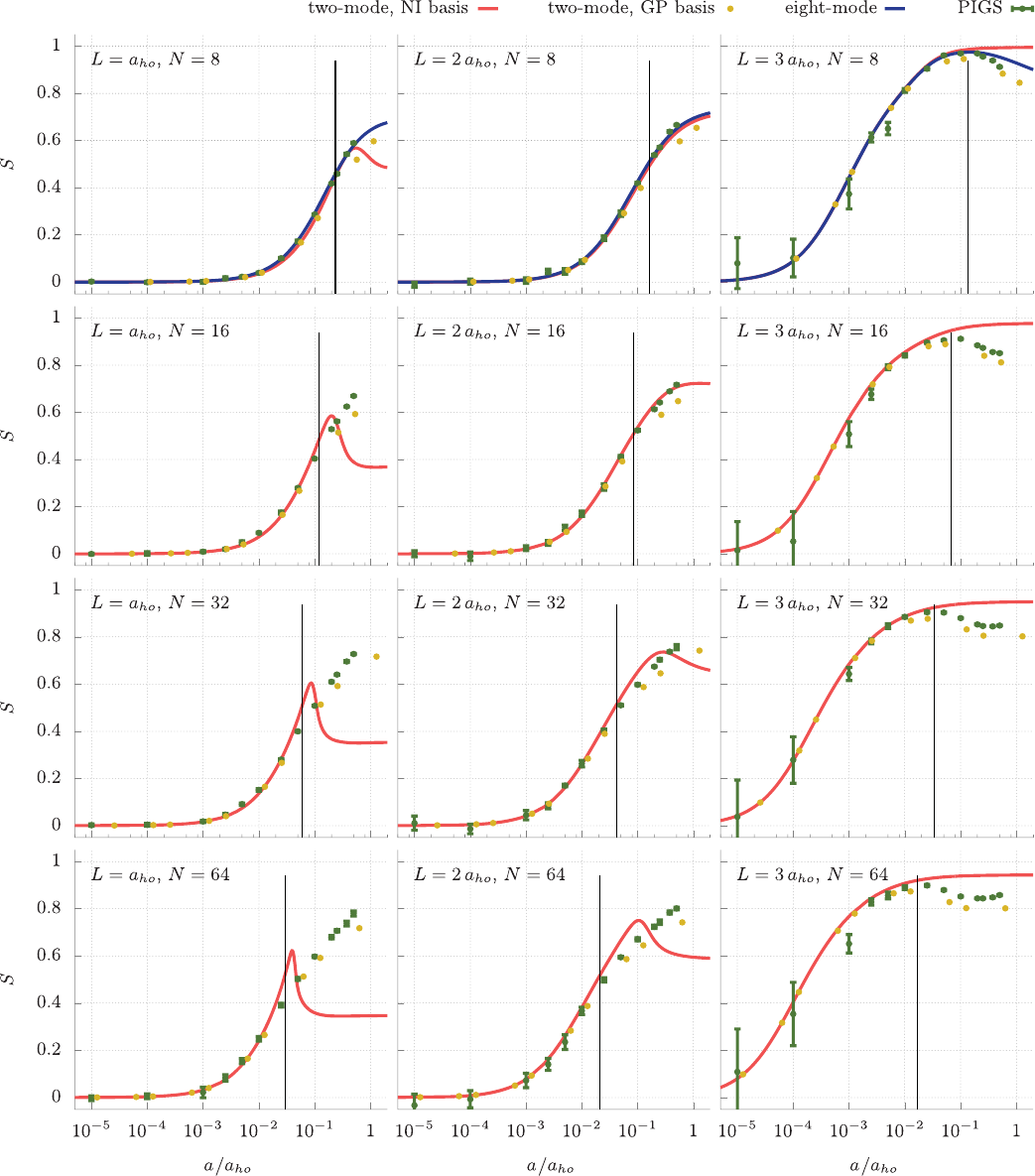}
\caption{\label{fig:PIGSsqueezing}(Color online) Squeezing $S$ vs.~scattering length $a$ for four
different particle numbers ($N=8$, 16, 32, and 64, from top to bottom) and three different
potentials ($L=a_{ho}$, $2\,a_{ho}$, and $3\,a_{ho}$, from left to right).  The plots include
results from the two-mode non-interacting model, the two-mode GP model, the eight-mode
non-interacting model, and the PIGS simulations. Vertical lines indicate the value of $a=a_{NI}$
below which we expect the two-mode models in the NI basis to be valid (Eq.~\eqref{eq:PIGSbreakdownNI}).}
\end{figure*}

Fig.~\ref{fig:PIGSsqueezing} shows the squeezing parameter $S$ as a function of $a$
for the PIGS calculations, together with comparisons to the corresponding two-mode and
eight-mode model results.  For the two-mode results, we include calculations made with the
non-interacting (NI) basis as well as a number of calculations made with the computationally more
expensive Gross-Pitaevskii (GP) basis, while for eight-mode results all calculations are made with
the NI basis.

By analogy to~\cite{Milburn1997}, we compute a criterion of validity for the two-mode model
in the NI basis, $a \ll a_{NI}$, where
\begin{equation}
a_{NI} = \frac{1}{N}\sqrt{\frac{9\pi}{8L}}\, .
\label{eq:PIGSbreakdownNI}
\end{equation}
Values of $a_{NI}$ are indicated in Fig.~\ref{fig:PIGSsqueezing}.  In general, we see that
both the two- and eight-mode models agree well with the PIGS results when this condition is met
but both deviate, to different extents, as $a$ increases beyond $a_{NI}$. As expected,
the two-mode, non-interacting model shows the first deviation from the PIGS results, while
the two-mode GP model shows agreement within a wider range of $a$ values.  In general,
the two-mode GP solutions mimic the qualitative behavior of the PIGS solutions, although
they systematically underestimate the amount of squeezing found in the PIGS results
and increasingly diverge from the PIGS solutions as $a$ increases. For the PIGS results
with $a=0.5$, this difference ranges in value from $\Delta S = 0.02$ to 0.1 depending
on the value of $L$ and $N$. The eight-mode model (only computed for $N=8$,
in the top panels of Fig.~\ref{fig:PIGSsqueezing}) shows the best overall agreement
with the PIGS results of the three finite basis calculations, including the non-monotonic
dependence on $a$ for barrier height $L=3$ (top right panel). Nevertheless, the eight-mode model
results also deviate from the PIGS results for sufficiently large $a$, and they
are only computable for small values of $N$.

This deviation between the finite basis models and the PIGS results is quite generic.  All such
expansions, regardless of the basis chosen, will eventually fail as both the intrinsic interaction
strength $a$ and the number of particles $N$ increases, since the particles increasingly populate
more delocalized states with population on both sides of the barrier, as we discuss in detail below.
In contrast, the PIGS approach allows arbitrary configurations of particles and may thus be regarded
as an ``infinite-mode'' model that can describe the ground state for arbitrary values of $a$ and $N$.

As mentioned above, the GP two-mode model systematically underestimates the PIGS squeezing results.
In contrast, whether the non-interacting models tend to over- or underestimate the PIGS results
for large $a$ depends on the strength of the double well barrier (i.e., the value of $L$).
For the non-interacting models with strong barriers
($L=3\,a_{ho}$), we find less squeezing the more accurate the model (i.e., when going from NI
two-mode to eight-mode to PIGS), whereas for weak barriers ($L=a_{ho}$), we find the opposite.
The intermediate-strength case ($L=2\,a_{ho}$) is a ``crossover'' between the other two cases,
where the non-interacting, finite basis models and PIGS calculations agree more closely.

In Sec.~\ref{sec:numdist}, we discussed a ``conventional'' mechanism for why one would expect
squeezing to increase with intrinsic interaction strength.  As $a$ increases, the ground state
changes to minimize the interparticle interaction energy, disfavoring configurations with many
particles on the same side of the double well.  Thus, strong repulsive interactions should suppress
number fluctuations and therefore increase squeezing.  However, this effect is insufficient
to explain the richer squeezing behavior we see in Fig.~\ref{fig:PIGSsqueezing},
in particular, the lack of monotonicity and the significant dependence on the barrier height
of the double well. We propose two additional mechanisms to account for this behavior.

\subparagraph{Delocalization Mechanism}

Like the conventional mechanism, the delocalization mechanism also involves changes
in the ground state that minimize interaction energies; it is suggested by the behavior
of the eight-mode ground state. As discussed in Sec.~\ref{sec:8moderesults}, the eight-mode ground
state contains increasingly large occupation of modes in the $n=2$ energy level as $a$ increases.
If we treat PIGS conceptually as an ``infinite-mode'' model, then we would expect similar behavior
in our simulations (i.e., occupation of $n=2$ and higher modes).

The more that a set of modes is delocalized into the ``wrong'' side of the double well,
the less squeezing it can support.  Certain modes with higher $n$ tend to be delocalized more than
the $n=1$ modes (the only ones present in the two-mode model), which implies that they can support
less squeezing.  Hence, when strong repulsion drives particles into modes of higher $n$,
it is driving some of them into modes that support less squeezing.  Thus, this mechanism
produces opposite results to the conventional mechanism, where increased $a$ leads to increased $S$.
Both of these mechanisms are independent of the shape of the double well (i.e., of $L$).

The delocalization mechanism is also relevant in understanding the differences between using
Gross-Pitaevskii basis states rather than non-interacting basis states.  Because the GP equation
contains a repulsive term, for a given double well geometry and value of interaction $a$,
the GP states are even broader, i.e., more delocalized, than NI states.  Hence GP states will tend
to support less squeezing than the corresponding NI states, as evident
in Fig.~\ref{fig:PIGSsqueezing} at larger $a$ values.

\subparagraph{Tunneling Mechanism}

The tunneling mechanism can be understood in the finite basis representation as a result
of the presence of higher modes leading to an increase in the number of types
of two-body tunneling terms present in the Hamiltonian.  As
discussed in detail in Sec.~\ref{subsec:2moderesultssqueezing}, two-body tunneling terms dominate the dynamics of the system when both $\chi_{1}$ and $\chi_{2}$ are large,
which occurs when the interaction parameter $a$ is large and the potential barrier parameter $L$ small.
In the two-mode model with the NI basis, these tunneling
terms force the ground state to occupy even-numbered Fock states only, causing large
number fluctuations and hence little squeezing.
Marked visual evidence of this is found in the striping of the distribution of Fock state components of the ground state (Figs.~\ref{fig:groundstateVSa}, \ref{fig:chi2plot}, and \ref{fig:chi1plots}).

The availability of higher modes
changes this situation by dramatically increasing the variety of two-body tunneling terms
in the Hamiltonian: two particles can tunnel from any two modes to any other two modes, as long
as the total value of the $z$-component of their angular momentum is conserved.
Hence, the alternating-Fock-state restriction is lifted and there is no striping seen for the eight-mode model in Fig.~\ref{fig:8modegroundstateVSa}.   Since the ground state can now have contributions
from Fock states with any value of $|n_l - n_r|$, there is a reduction in the occupation
of modes with large values of $|n_l - n_r|$, and hence squeezing will be greater than predicted
by the two-mode model and will increase as modes are further added. Again treating the PIGS simulations
as ``infinite-mode'' implies that this effect will be even more pronounced in the Quantum Monte Carlo
results than in the eight-mode model results.

This tunneling mechanism varies in importance depending on the value of $L$.  For small $L$,
it operates as described above.  However, for large $L$, tunneling between the wells is highly
suppressed for all models ($\chi_{2}$ is small).  In the two-mode case, this drives the ground state
towards $\ket{N/2}$, an equal splitting of particles.  The presence of higher modes has little
effect on this distribution, again because all tunneling between the wells is suppressed.
Therefore, the amount of squeezing is not affected for large $L$.

The tunneling mechanism also varies in importance depending on whether NI or GP basis
states are employed in the calculations.  For GP states, we observe that the value of
the two-body tunneling parameter $\chi_{2}$ is small for large $a$, regardless of the value of $L$.  Hence, two-body tunneling is never relevant when a basis
of GP states is used, and therefore squeezing in the GP two-mode model is not suppressed
at large $a$ values for small $L$  as it is with the NI two-mode model (see left panel
of Fig.~\ref{fig:PIGSsqueezing}).

\begin{figure*}
\includegraphics{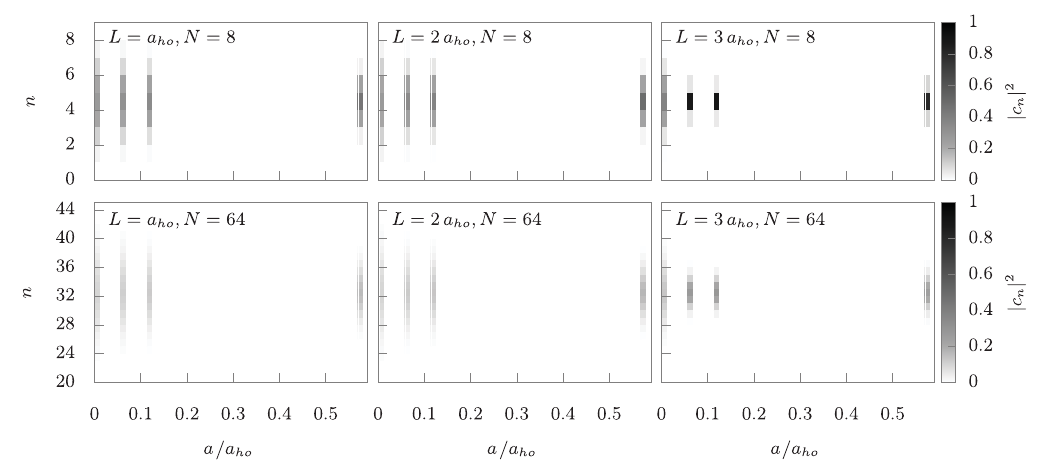}
\caption{\label{fig:groundstateVSaFORgp}
The components of the ground state $|c_{n}|^2 = |\braket{n|\psi_{ground}}|^2$ for the double well
potential as a function of scattering length $a$, as computed in the
exact two-mode model using a GP basis,
for $N=8$ (top row) and $N=64$ (bottom row), with $L=a_{ho}$ (left panel), $2\,a_{ho}$ (center panel), and $3\,a_{ho}$ (right panel). The components are computed only for the values of $a$ corresponding
to the GP solutions in Fig.~\ref{fig:PIGSsqueezing}. There is no evidence here of the striping
as a function of $n$ that is seen for calculations with the NI basis in Fig.~\ref{fig:groundstateVSa}.}
\end{figure*}

Another way to confirm the effects of the two-body role in the tunneling mechanism is through
the structure of the ground state. We have already noted the comparison between the striped ground state distributions of the two-mode models (Figs.~\ref{fig:groundstateVSa}, \ref{fig:chi2plot}, and \ref{fig:chi1plots}) and  the non-striped distributions of the eight-state models (Fig.~\ref{fig:8modegroundstateVSa}). In addition, Fig.~\ref{fig:groundstateVSaFORgp} shows
the components of the ground state for the GP two-mode model.  As opposed to its counterpart in the NI basis (Fig.~\ref{fig:groundstateVSa}), Fig.~\ref{fig:groundstateVSaFORgp} shows no striping, i.e., no alternation between even- and odd-numbered Fock states. This pattern is consistent with an insignificant amount of two-body tunneling deriving from the smaller value of $\chi_2$ associated with GP states.

\subparagraph{Interaction of Mechanisms}

In summary, when considering models using the NI basis, the delocalization mechanism tends
to reduce the amount of squeezing as the number of modes increases and is relevant for any $L$,
while the tunneling mechanism tends to increase squeezing as the number of modes increases
and is only important for small $L$.  The combination of these effects explains the patterns seen
in Fig.~\ref{fig:PIGSsqueezing} among the NI two-mode, eight-mode, and PIGS results.
For small $L$, the tunneling mechanism dominates, and there is more squeezing than predicted
by the two-mode model.  For large $L$, the delocalization mechanism dominates, and there is
less squeezing than predicted by the two-mode model.

The GP two-mode results deviate from this pattern.  In general, the
spatial broadening, i.e., the greater delocalization of the GP states
relative to NI states that results from the repulsive term in the GP equation, gives rise to a systematic reduction in squeezing for the GP two-mode results relative to the other results. This includes those
of the PIGS calculations and thus confirms the significance of the delocalization mechanism.
The only situation where the GP squeezing is not the smallest is for the smaller $L$ values,
where the NI two-mode squeezing is the lowest value (left column and bottom two panels
of the middle column of Fig.~\ref{fig:PIGSsqueezing}). In that situation, the NI two-mode squeezing
is dominated by the tunneling mechanism, which leads to low squeezing at small $L$ values as described
in detail above.

A marked feature of the Quantum Monte Carlo results is the lack of monotonicity in the $L=3\,a_{ho}$
PIGS data (right-hand column of Fig.~\ref{fig:PIGSsqueezing}).  This effect reflects an interplay
between the ``conventional'' mechanism, which increases squeezing, and the delocalization mechanism,
which suppresses squeezing, as the interactions become increasingly repulsive.

\subsubsection{Fragmentation and Depletion}
\begin{figure*}
\includegraphics{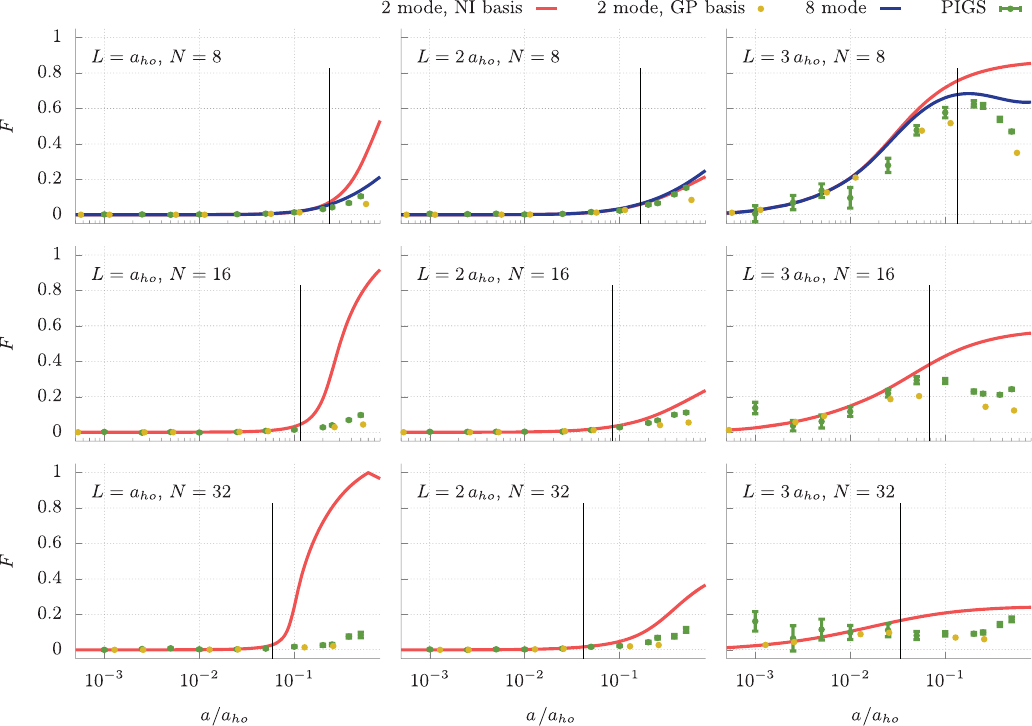}
\caption{(Color online) Fragmentation $F$ vs.~scattering length $a$ for three
different particle numbers ($N=8$, 16, and 32, from top to bottom) and three different
potentials ($L=a_{ho}$, $2\,a_{ho}$, and $3\,a_{ho}$, from left to right).  The plots include
results from the two-mode non-interacting model, the two-mode GP model, the eight-mode
non-interacting model, and the PIGS simulations. Vertical lines indicate the value of $a=a_{NI}$
below which we expect the two-mode models in the NI basis to be valid (Eq.~\eqref{eq:PIGSbreakdownNI}).}
\label{fig:PIGSfragmentation}
\end{figure*}

We now discuss the fragmentation and depletion results. Fig.~\ref{fig:PIGSfragmentation}
shows the fragmentation parameter $F$ as a function of $a$ for PIGS simulations with the same
ranges of $N$, $L$, and $\alpha$ as before, together with comparison to the NI two-, GP two-,
and NI eight-mode models.  For weak barriers ($L = 2\,a_{ho}$ and $3\,a_{ho}$), there is
a modest amount of fragmentation at large $a$, whereas for $L = 3\,a_{ho}$ and small $N$,
there is a much larger amount of fragmentation for large $a$.  The amount of fragmentation
decreases with increasing $N$.  Additionally, as for squeezing, we find that the amount
of fragmentation does not vary monotonically with $a$.

As before, the eight-mode model and the GP two-mode model generally show better agreement
with the PIGS simulations at larger $a$ values than the non-interacting two-mode model,
in particular showing a maximum and non-monotonic behavior for $L=3$.  The vertical lines
in Fig.~\ref{fig:PIGSfragmentation} again indicate the values of $a$ below which the two-mode model
is expected to be sufficient to describe the physics of the system ($a \leq a_{NI}$).
As was seen for squeezing, the truncated basis models and PIGS results agree well when $a < a_{NI}$,
but deviate as $a$ increases beyond this value.  In particular, the GP two-mode results tend
to underestimate the amount of fragmentation given by the PIGS simulations, while the NI two-mode
and eight-mode results tend to overestimate them.

These trends can be accounted for by the mechanisms described in
Sec.~\ref{subsub:QMCsqueezingresults}.  Fragmentation in the double well system increases when
the wells are more ``isolated'', i.e., when it is unlikely for a particle to tunnel from one well
to the other.  Thus, potentials with larger $L$ experience more fragmentation, as we see
in the PIGS results (as well as the eight-mode and GP two-mode results).  The delocalization
mechanism leads to the occupation of modes with higher values of $n$, some of which have more
spillover into the ``wrong'' side of the double well than the $n=1$ modes.  Hence, the occupation
of these modes will lessen the isolation of the wells, and hence reduce fragmentation,
by putting particles in states that span both sides of the double well barrier.  Thus, we expect
(and find) that the PIGS results will have less fragmentation than the eight-mode results,
which in turn will have less fragmentation than the NI two-mode results.  As discussed above,
the GP states are naturally ``delocalized'' relative to the NI states, so the GP two-mode model
shows less fragmentation than the NI two-mode model; in fact, it shows the least fragmentation
of the three models.

The tunneling mechanism helps to explain the anomalously high fragmentation seen in the NI two-mode
model data for small $L$ ($L=1$).  Recall that in the two-mode model, fragmentation $F$
grows as the product of the occupation of adjacent Fock states ($c_{n}c_{n+1}$) decreases.
For the NI two-mode model, two-body tunneling dominates for small $L$, causing a striped ground
state and consequently large fragmentation.  However, for the GP two-mode model, tunneling
is suppressed (the $\chi_{2}$ parameter is small), which reduces striping and thus reduces
fragmentation.  Similarly, the presence of additional two-body tunneling terms for the
eight-mode model and the ``infinite-mode'' PIGS simulation lead to a ground state with
only modest amounts of fragmentation.

\begin{figure*}
\includegraphics{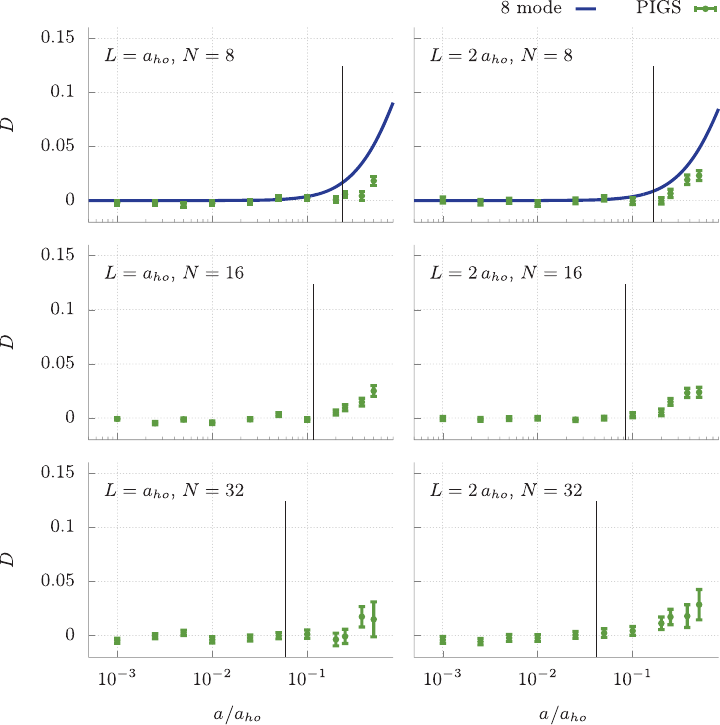}
\caption{(Color online) Depletion $D$ vs.~scattering length $a$ for three
different particle numbers ($N=8$, 16, and 32, from top to bottom) and
two different potentials (left: $L=a_{ho}$, right: $L=2\,a_{ho}$).
The plots include results from the the eight-mode non-interacting model (blue lines) and
the PIGS simulations (green circles, with error bars); the two-mode models are excluded because
for these $D=0$ by definition.  The vertical black lines indicate the value of $a=a_{NI}$ below
which we expect the models to be valid (Eq.~\eqref{eq:PIGSbreakdownNI}).}
\label{fig:PIGSdepletion}
\end{figure*}

Finally, we comment on the degree of depletion seen in the PIGS data.  Fig.~\ref{fig:PIGSdepletion}
shows the depletion parameter $D$ as a function of $a$ for both the eight-mode model
and the PIGS simulations, with the $L=1$ and $L=2$ potentials.  recall that we define $1- D$
as the combined population of the two natural orbitals with highest occupancy
(see Eq.~\eqref{eq:D}), so for the two-mode model $D$ is zero by definition.
For the PIGS simulations at large $a$ ($>0.1$), we see a modest nonzero amount of depletion
that is of comparable magnitude in all calculations. From the analysis
of Bogoliubov~\cite{Bogoliubov1947}, we know that depletion begins to become significant
in a homogeneous BEC when  $an^{1/3}$ approaches 1, (where $n$ is the particle density),
i.e., when the gas is no longer dilute. We can estimate the relevant density $n$
by the maximum value of the quantity $N\rho_{x}(x)\rho_{y}(y)\rho_{z}(z)$, with $\rho_{i}$
the one-body density in the $i$ direction, since depletion will be dominated by the parts
of the BEC with greatest density. We find that for the PIGS data, $an^{1/3}$ is a linear function
of the scattering length $a$ that is essentially independent of $N$ and $L$ and that reaches
values of order ~0.9 for $a>0.1$, consistent with the PIGS depletion results shown
in Fig.~\ref{fig:PIGSdepletion}.

\subsubsection{Universal Scaling of Squeezing with $Na$}
\begin{figure*}
\includegraphics{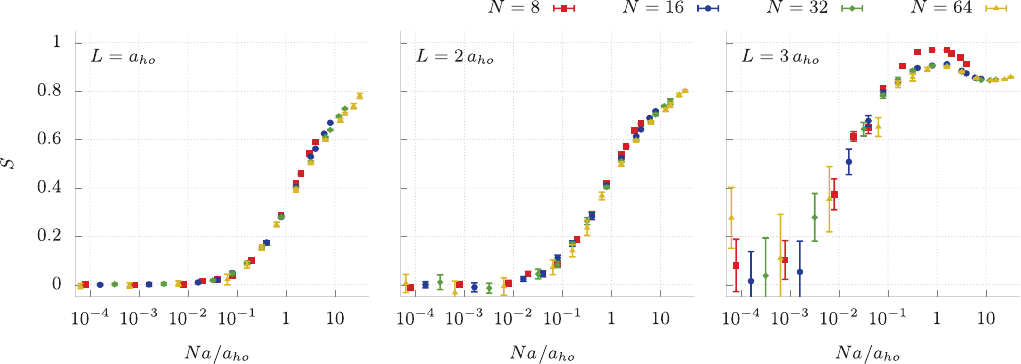}
\caption{(Color online)  Universal scaling of the Quantum Monte Carlo (PIGS) results
for the squeezing parameter $S$ vs.~$Na$ (the product of the number of particles and the scattering
length) for three different potentials ($L=a_{ho}$, $2\,a_{ho}$, and $3\,a_{ho}$, from
left to right).}
\label{fig:PIGSsqueezing_Na}
\end{figure*}

Fig.~\ref{fig:PIGSsqueezing_Na} shows the PIGS squeezing data
of Fig.~\ref{fig:PIGSsqueezing}, plotted now as a function of $Na$ for each value of $L$.
The most striking feature of this plot is that the data for the various values of $N$ overlap
each other (with one exception, discussed below). The same universality is found when plotting
versus $(N-1)a$, indicating that the scaling is not dependent on the range of $N$ employed here.
Thus, we have found that, to a good approximation, $S$ is a universal function of the product
$Na$ for the potentials and ranges of $N$ and $a$ presented here. The one exception is the $N=8$
data for the highest barrier ($L=3\,a_{ho}$), which shows slightly more squeezing at large $Na$
than seen with the larger values of $N$ (see discussion below).

This universal scaling of squeezing with $Na$ across all of our data implies
that we can apply our squeezing results to systems with larger values of $N$ than
simulated here, for correspondingly lower values of $a$. Since the largest value of $Na$ in our
data set is $32$ ($N=64, a=0.5$), the interesting non-monotonicity seen for $Na$ values greater
than unity is thus directly relevant to current experiments with, e.g., $N\sim 10^{3}-10^{5}$
and $a/a_{ho}\sim 10^{-3}$, or even larger numbers of atoms $N\sim 10^{6}-10^{7}$~\cite{Streed2006}
when the interaction is tuned to smaller values $a/a_{ho}\sim 10^{-5}$ by exploiting Feshbach
resonances~\cite{Pollack2009}.  Such universal scaling also suggests that to compute additional
results for larger $N$, one could instead simulate systems with the same $N$ as employed here
but larger $a$ values. This key observation therefore mitigates the system size limitation
inherent in the Quantum Monte Carlo calculations for as long as this universality continues
to hold (i.e., as $N$ and $a$ increase further).

The universal scaling of squeezing with $Na$ (or equivalently, $Na/a_{ho}$, since
our scattering length is scaled by $a_{ho}$) resembles the scaling of simple mean field estimates
of the ratio of interaction energy to kinetic energy for dilute trapped gases~\cite{Dalfovo1999}.
The same scaling is also found for the PIGS ground state energy~\cite[Fig.~6.2]{Corbo2013},
consistent with the ratio of interaction to kinetic energy. For $Na>1$, the PIGS results
for both the ground state energy and squeezing are thus generally consistent with
the interaction energy dominating over the kinetic energy of particles in the double well system.
The deviation for $N=8, L=3$ noted above may then be understood as arising because this is
the set of parameters with both the highest zero point energy and the smallest number of particles.
This results in a far greater relative contribution of the kinetic energy and therefore
a significant deviation away from the parameter regime where the interaction energy dominates
the physics of the system.

\begin{figure*}
\includegraphics{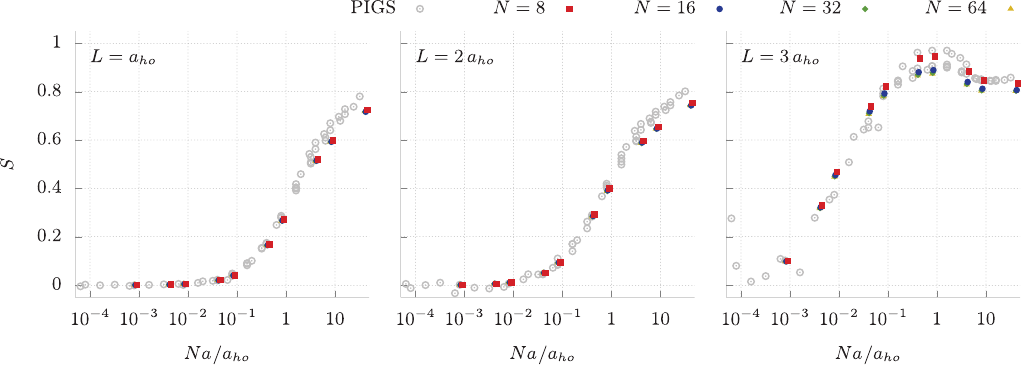}
\caption{(Color online)  Universal scaling of the results of the two-mode GP model
for the squeezing parameter $S$ vs.~$N a$ (the product of the number of particles and the scattering
length) for three different potentials ($L=a_{ho}$, $2\,a_{ho}$, and $3\,a_{ho}$, from
left to right).  The gray dots are the PIGS data from Fig.~\ref{fig:PIGSsqueezing_Na} for
comparison, with error bars removed.}
\label{fig:GPsqueezing_Na-withPIGS}
\end{figure*}

We note that ground state mean field estimates obtained from the two-mode calculations
with GP basis functions are typically also expected to scale with $Na$ (or $(N-1)a$
for small $N$) because the effective interaction strength scales with $Na$~\cite{Dalfovo1999}.
Indeed, the GP results in Fig.~\ref{fig:PIGSsqueezing} also show such scaling
(see Fig.~\ref{fig:GPsqueezing_Na-withPIGS}) and also show non-monotonic squeezing behavior
for $Na>1$. However, there are nevertheless differences between the PIGS and GP scaling functions
that increase with $Na$, as is evident from Fig.~\ref{fig:GPsqueezing_Na-withPIGS},
where the scaling function for the two-mode GP results is seen to be systematically lower
than that for the PIGS results for the larger $Na$ values.  This deviation from the PIGS results
can be understood as reflecting the inaccuracies of the GP basis in representing both
delocalization across the barrier and two-body tunneling (see discussion above),
as well as the inevitable break down of a two-mode description at large enough $Na$ values.

In contrast to the universal scaling behavior of squeezing, the fragmentation $F$ shows
no universal scaling with $Na$ (or $(N-1)a$ for small $N$) other than in the regimes where
$F\sim 0$.  This may be understood in terms of the different natures of fragmentation and squeezing.
While squeezing is a real space property that integrates over the behavior of all of
the natural orbitals, fragmentation is determined by the behavior of specific natural orbitals
and will be more sensitive to the geometry of the potential.  It is thus not surprising that
this property does not scale generically with a ratio of total interaction energy
to kinetic energy.

\section{Summary and Conclusions}
\label{sec:conclusion}
We have presented a detailed analytical and numerical study of the squeezing and
fragmentation exhibited by the ground state of an ultracold, bosonic gas
in a three-dimensional double well trap for a variety of particle number, interaction strengths,
and double well trap geometries, using exact ground state calculations with Quantum Monte Carlo
methods and comparing to truncated basis models with two and eight modes only.
In making these comparison, we have extended the previous two-mode analyses with analysis
of the exact two-mode Hamiltonian and investigated for the first time the squeezing
and fragmentation phenomena of a recently-proposed eight-mode model.  Using numerically exact
Quantum Monte Carlo methods to simulate the system led to a number of interesting and surprising
discoveries about these well-known systems, most notably the fact that squeezing and fragmentation
show non-monotonic behavior with intrinsic interaction strength
$a$, particularly for large barrier heights $L$.

The quantitative understanding gained from this study allows for a more sophisticated qualitative
picture of the way in which squeezing and fragmentation come about in a double well system
than was previously possible. Recall that the suppression of number fluctuations corresponds
to increased squeezing and the suppression of tunneling corresponds to increased fragmentation.
Both the old and the new qualitative pictures start the same way:
\begin{quote}
The ground state of the noninteracting double well is a product of the one-body ground state
of each particle, and these one-body states each occupy both wells equally.  Hence, if one
were to measure the number of particles in the left well minus the number in the right well,
one could get any value from $N$ to $-N$.  Therefore, number fluctuations are large and squeezing
is small.  An equivalent way to think of this situation is that the structure of the
noninteracting ground state is such that tunneling is strong, and therefore fragmentation is
small.  In fact, both $S$ and $F$ are defined to be 0 in the noninteracting case.
\end{quote}
In the old picture, the introduction of repulsive interacting proceeds like this:
\begin{quote}
The introduction of repulsive interactions causes the system to minimize its interacting energy by
suppressing configurations in which many particles are in one well and few are on the other.  This
reduces number fluctuations and increases squeezing.  Additionally, moving towards a configuration
in which $N/2$ particles are locked into each side of the double well suppresses tunneling
and increases fragmentation.  These effects increase with increasing interacting strength.
\end{quote}
However, in the new picture, the introduction of repulsive interactions proceeds as follows:
\begin{quote}
The introduction of repulsive interactions causes the system to minimize its interacting energy
in several ways.  One way is to suppress configurations in which many particles
are in one well and few are on the other. This increases squeezing and fragmentation as
in the old picture.  However, the system can also reduce its interaction energy by promoting
particles to modes in higher energy levels ($n > 1$), leading to increased contributions
from tunneling and delocalization.  One can approximate these modes as each
being localized in one of the two wells, although they will extend into the ``wrong'' well based
on the strength of the double well barrier.
There are some modes in each energy band $n>1$
that extend into the wrong well much further than the $n=1$ modes.  Hence, a ground state
dominated by modes with larger values of $n$ naturally have larger number fluctuations and
tunneling than ground states dominated by $n=1$ modes, and therefore they exhibit less squeezing
and fragmentation.  These two effects compete with each other to determine the overall amount
of squeezing and fragmentation, which is not monotonic in many cases.
\end{quote}
The above description implicitly relies on the language of expansions in one-body bases.
However, such descriptions require truncation to finite bases for calculations,
and as $a$ increases, any finite basis truncation will eventually fail. In this work,
we have seen explicitly how the eight-mode model with non-interacting basis states extends
the validity of a finite basis description to larger $a$ values that that of the conventional
two-mode, non-interacting model, but we also saw that the eight-mode model does not describe
squeezing or fragmentation accurately for large values of $Na$, in particular for values relevant
to current experiments.  Similarly, we also saw that the two-mode model produces valid results
for larger values of $a$ when using a basis constructed from solutions to the Gross-Pitaevskii
equation rather than a non-interacting basis; however, both of these are less valid for large
$Na$ than the eight-mode model.

Thus, for large values of $a$ and/or $N$, our results show  that one cannot appeal
to either mean-field or multi-mode
descriptions to correctly predict the amount of squeezing and fragmentation exhibited
by the system. Instead, one must deploy the full machinery of a numerically-exact method such as
Quantum Monte Carlo, which allows for full three-dimensional calculations without restriction
to a truncated basis set representation. The PIGS Quantum Monte Carlo results presented here
show that for a given double well potential, the amount of squeezing is a function solely
of the product $Na$. This universal scaling implies that our results for
$Na = 1 - 64$, corresponding to, e.g., $N=10^3-10^5$ and $a\sim 10^{-3}$, lie within the regime
of current experiments with cold atoms trapped in double well potentials.
In contrast, the fragmentation shows no such universal scaling.

This study also showed that the potential barrier height, parameterized in this work by $L$,
and the intrinsic interaction strength, parameterized by the $s$-wave scattering length $a$,
are independent parameters that control the number squeezing, fragmentation, and
depletion differently.  Thus, it is important to study the dependence on each of these
independently.  In addition, it is evident that given the non-monotonic behavior
of the number squeezing at large values of $Na$, experimental verification of this behavior
will require studies for which that product can be carefully controlled. This suggests experimental studies that hold constant the total number of particles $N$,
which requires double well realizations having high trapping potentials, regardless of the internal barrier height.  On the theoretical side, the present studies have shown the potential usefulness
of a truncated basis calculation using mean field basis states from GP solutions,
rather than non-interacting basis functions.  While there are significant computational challenges
in extending this approach to fully self-consistent sets of GP basis functions, the agreement
with PIGS results for squeezing and fragmentation at all except the largest values of $N$ and $a$
indicate that this might be a useful avenue for further numerical studies
in the strongly interacting regime.

Finally, we would like to revisit one of the main motivations of this work discussed
in the introduction, namely the application of squeezed states to reduce the measurement
uncertainty of atom interferometers.
One way to generate a highly squeezed state is to use a Feshbach resonance to tune the intrinsic
interaction strength of the atoms in a BEC, thereby changing the amount of squeezing exhibited
by the system~\cite{Pollack2009}.  However, what interaction strength is the one that
\emph{maximizes} squeezing?  In the context of the old qualitative picture,
which is based on the nearly-degenerate two-mode description, the answer is simple:
stronger repulsive interactions mean more squeezing, so one should tune $a$ to as large a value
as possible.  However, we have shown that the real picture is far more complicated.
In particular, in many situations squeezing does not increase monotonically with interaction
strength, and one has to consider the contributions from delocalization and tunneling
in addition to inter-particle interactions in order to understand the detailed behavior. In
these situations, there is an optimal value of $a$ that maximizes squeezing.
For large $N$, this optimal value cannot be predicted through the use of $n$-mode models,
but must instead be calculated from the full Hamiltonian with an exact
but computationally expensive method, such as the path integral ground state (PIGS)
Quantum Monte Carlo approach that was employed here. With the increasingly rapid advances
in experimental methods for the study of Bose-Einstein condensates, we look forward to laboratory
confirmation of the results of this study in the near future.

\section{Acknowledgements}
\label{sec:acknowledgements}

This research was supported by the National Science Foundation and by the UC Lab Fees Research
Program under a grant to the University of California, Berkeley and Lawrence Livermore National
Laboratory.  J.C.C.~and K.B.W.~were also supported by funding from the National Science Foundation
Grant No. CHE-1213141. J.L.D.~acknowledges work performed under the auspices of the U.S. Department
of Energy by Lawrence Livermore National Laboratory under Contract No. DE-AC52-07NA27344.
K.B.W.~and J.C.C.~also thank the Kavli Institute for Theoretical Physics for their hospitality
and for supporting this research in part by the National Science Foundation Grant No. PHY-1125915.

\bibstyle{apsrev4-1}
\bibliography{DoubleWell}
\end{document}